\documentclass[lettersize,journal]{IEEEtran}
\usepackage{graphicx}
\usepackage[caption=false,font=normalsize,labelfont=sf,textfont=sf]{subfig}
\usepackage{array}
\usepackage{booktabs}
\usepackage{multirow}
\usepackage{algorithm}
\usepackage{algorithmic}
\usepackage{amsmath,amsfonts}
\usepackage{bm}
\usepackage{cite}

\begin{document}

\title{MS-HLMO: Multi-scale Histogram of Local Main Orientation for Remote Sensing Image Registration}

\author{Chenzhong Gao, Wei Li, \emph{Senior Member, IEEE}, Ran Tao, \emph{Senior Member, IEEE}, Qian Du, \emph{Fellow IEEE}
\thanks{This work is supported by the National Key R\&D Program of China (Grant No. 2021YFB3900502) and National Natural Science Foundation of China (61922013). (Corresponding author: Wei Li; e-mail: liwei089@ieee.org)}

\thanks{C. Gao, W. Li, R. Tao are with the School of Information and Electronics, Beijing Institute Technology, Beijing 100081 China; Beijing Key Laboratory of Fractional Signals and Systems, Beijing Institute of Technology, Beijing 100081, China (e-mail: gao-pingqi@qq.com).}

\thanks{Q. Du is with the Department of Electrical and Computer Engineering, Mississippi State University, Mississippi State, MS 39762, USA (e-mail: du@ece.msstate.edu).}
}


\maketitle

\begin{abstract}
Multi-source image registration is challenging due to intensity, rotation, and scale differences among the images. Considering the characteristics and differences of multi-source remote sensing images, a feature-based registration algorithm named Multi-scale Histogram of Local Main Orientation (MS-HLMO) is proposed. Harris corner detection is first adopted to generate feature points. The HLMO feature of each Harris feature point is extracted on a Partial Main Orientation Map (PMOM) with a Generalized Gradient Location and Orientation Histogram-like (GGLOH) feature descriptor, which provides high intensity, rotation, and scale invariance. The feature points are matched through a multi-scale matching strategy. Comprehensive experiments on 17 multi-source remote sensing scenes demonstrate that the proposed MS-HLMO and its simplified version MS-HLMO$^+$ outperform other competitive registration algorithms in terms of effectiveness and generalization.
\end{abstract}


\section{Introduction}
\label{sec:introduction}
Image registration is a key preprocessing step in remote sensing applications. With earth observation systems being developed rapidly in
recent years, to achieve in-depth ground analysis, the use of multi-source remote sensing images becomes popular in achieving in-depth ground analysis \cite{zhang2018feature,zhao2020joint,liu2020joint,zhang2020transfer,gao2021hyperspectral}. The accurate alignment of the multi-source images is a prerequisite of subsequent applications such as data fusion, change detection, joint analysis, and other techniques.

\begin{figure}[h!]
 \begin{center}
  \includegraphics[width=3.0in]{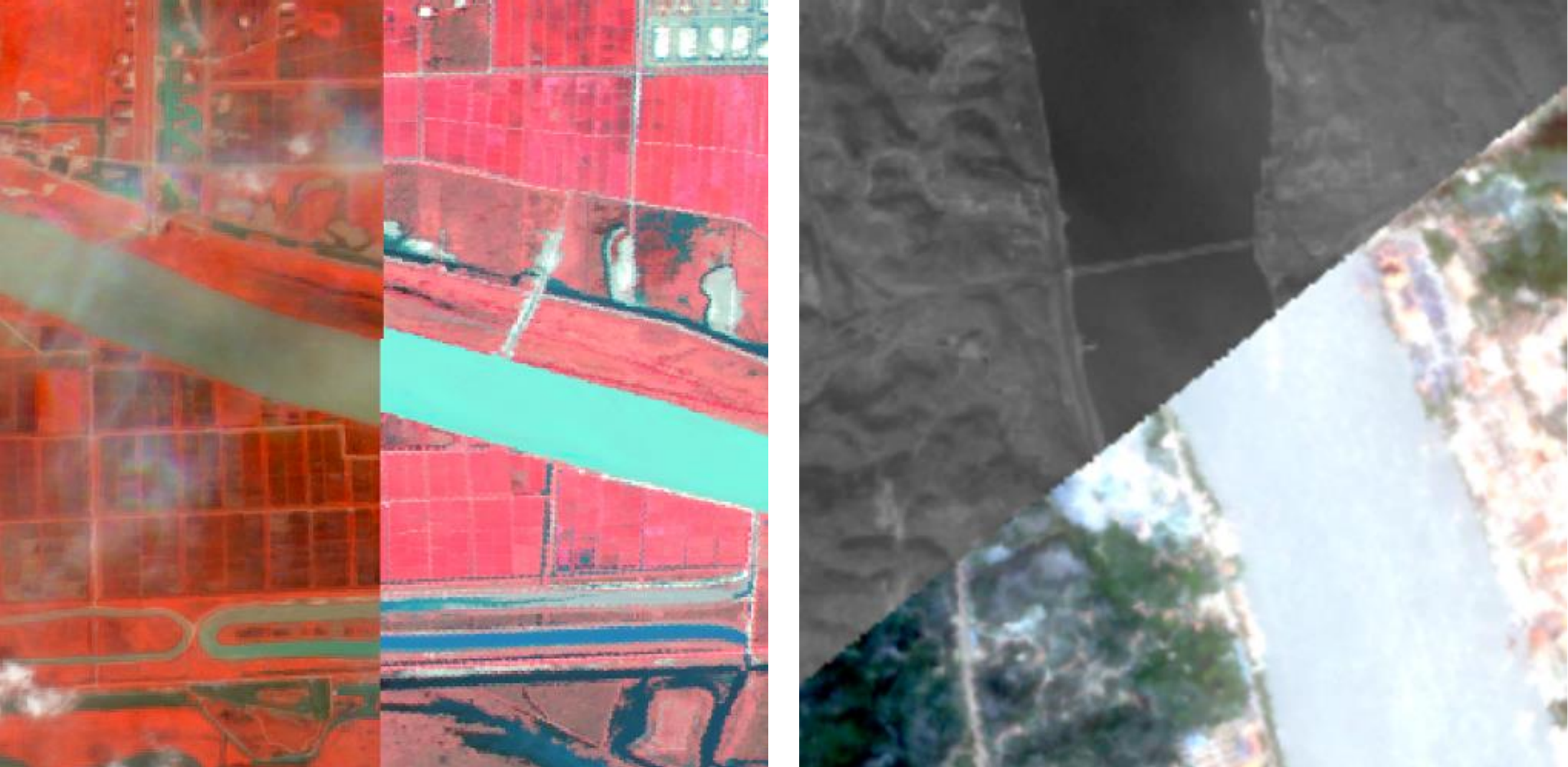}
  \caption{Examples of alignment failure of multi-source remote sensing images with only calibration information.}
  \label{fig:fail}
 \end{center}
\end{figure}

The purpose of image registration is to spatially align multiple images for the same scene. In general, remote sensing images are spatially corrected by reference coordinates (such as fictitious graticule) or image control points. However, there are inconsistent or missing spatial references. In particular, in multi-source remote sensing, the inconsistency of spatial correction between different sources is more obvious. This phenomenon can be seen from the examples shown in Fig.\ref{fig:fail}, which is about the correction of the multi-sensor image solely through the spatial reference information provided by the data source. It is obvious that the ground covers correspond inconsistently. In addition, for some UAV or ground-based images, there is no spatial reference information. It is very important to correct multi-source remote sensing images automatically by registration algorithms in order to provide accurately registered data for the same scene space.

Existing automatic image registration algorithms are generally divided into two categories, i.e., the traditional methods and the deep learning-based methods. Among them, the deep learning-based methods \cite{litjens2017survey} are relatively advanced, but a large amount of annotated samples are needed for model training, and the model is often highly targeted. The traditional ones are systematically classified into area-based and feature-based methods \cite{zitova2003image}. The area-based, also called intensity-based methods
register images by establishing the similarity measure of the intensity values or in a transform domain \cite{pratt1974correlation,de1987registration,viola1997alignment,le2002automated,pluim2003mutual,mahmood2011correlation,oliveira2014medical,tong2015novel}, and optimal geometric transformation parameters between the image pairs are found through optimization \cite{zhang2021multimodal}. Apart from efficiency and other factors, for multi-source images, this type algorithm is prone to fail when multi-modal differences are present.

Registration based on feature matching is a relatively mature technique, in which the transformation parameters are estimated through the coordinate correspondence of matched features. Scale-Invariant Feature Transform (SIFT) \cite{lowe1999object,lowe2001local,lowe2004distinctive} is one of the most classic, effective, and commonly used feature extraction and matching methods, and became the basis of many improved algorithms \cite{ke2004pca,mikolajczyk2005performance,bay2006surf,morel2009asift,sedaghat2011uniform,dellinger2014sar,sedaghat2015remote}. SIFT uses Gaussian scale-space, supported by scale-space theory \cite{lindeberg1994scale}, for feature points extraction, which also provides scale invariance. The feature points are then described by statistical local gradient information for points matching. This also lays the framework for the subsequent algorithms. In addition, many feature extraction algorithms have been proposed and applied to image registration \cite{moravec1980obstacle,harris1988combined,dalal2005histograms,rosten2008faster,calonder2011brief,rublee2011orb,leutenegger2011brisk,alahi2012freak,alcantarilla2012kaze,ye2019fast}.

However, most algorithms are only applicable to single-modal image registration. Algorithms for multi-modal image registration are more critical. Focusing on this, multi-modal image registration algorithms have been proposed. Chen et al. \cite{chen2010partial} proposed Partial Intensity Invariant Feature Descriptor (PIIFD) based on SIFT algorithm for multi-source retinal image registration, which overcame the problem of intensity difference problems. PSO-SIFT \cite{ma2016remote} presented a new gradient definition and an enhanced feature matching method for the registration of remote sensing images with intensity differences. Ye et al. developed Histogram of Orientated Phase Congruency (HOPC) \cite{ye2017robust} and Local HOPC (LHOPC) \cite{ye2018local} for multi-modal remote sensing registration, in which Minimum Moment of Phase Congruency (MMPC)-Lap is used for keypoints detection and an extended phase congruency model is used for feature description. Radiation-variation insensitive feature transform (RIFT) \cite{li2019rift} utilized a maximum index map (MIM) as the feature map, which is invariant to intensity diffeerence. The MIM is obtained by assigning the strongest response of the log-Gabor filtering at several predetermined orientations as the maximum index at each pixel. Multi-scale PIIFD (MS-PIIFD) \cite{2021Multi} improved PIIFD by adopting Gaussian scale-space, in which feature extraction and matching are completed in a multi-scale strategy. It achieves scale robustness and performs well on multi-modal images with scale differences.

The multi-modal differences of multi-source images result in multiple inconsistency of local features in the same area of image pairs, which makes it difficult to match correctly unless all inconsistencies are conquered. It is needed to handle the differences and enhance the similarity of the features. Therefore, through deep-going analysis of practical work and real data, we refine and model the differences of multi-source remote sensing images as
\begin{equation}
{{\bf{I}}_2} = {{\cal F}_{{\rm{Cut}}}}({{\cal F}_{{\rm{Tran}}}}({{\cal F}_{{\rm{Sc}}}}({{\cal F}_{{\rm{Rot}}}}({{\cal F}_{{\rm{Int}}}}({{\cal F}_{{\rm{Chg}}}}({{\bf{I}}_1}))))))
\end{equation}
where ${{\bf{I}}_1}$ and ${{\bf{I}}_2}$ are a multi-source image pair, ${{\cal F}_{{\rm{Chg}}}}( \bullet )$ represents the changes of the real ground object content sensed in the images, which is often caused by multi-temporal sensing.
${{\cal F}_{\rm{Int}}}( \bullet )$ denotes intensity transformation.
${{\cal F}_{{\rm{Rot}}}}( \bullet )$, ${{\cal F}_{{\rm{Sc}}}}( \bullet )$, and ${{\cal F}_{{\rm{Tran}}}}( \bullet )$ are spatial transformation of rotation, scale change, and translation, respectively, which are the basic operation of similarity transformation.
${{\cal F}_{{\rm{Cut}}}}( \bullet )$ represents the spatial cutting operation, which results in images covering different ranges of vision.
There may also be large perspective differences or severe local geometric distortions between images, which leads to more complex image distortion and registration modeling. Our goal is to first resolve the most common and fundamental problem of image differences. Among the above multi-modal differences, ${{\cal F}_{{\rm{Cut}}}}( \bullet )$, ${{\cal F}_{{\rm{Tran}}}}$, and ${{\cal F}_{{\rm{Chg}}}}( \bullet )$ have been solved by feature-based process. Thus, the primary goal is to find a feature robust to ${{\cal F}_{{\rm{Sc}}}}( \bullet )$, ${{\cal F}_{{\rm{Rot}}}}( \bullet )$, and ${{\cal F}_{\rm{Int}}}( \bullet )$ in multi-source images.

Based on the above analysis, a novel feature-based registration framework for multi-source remote sensing images is proposed through designing effective strategies for robust feature extraction that overcomes multiple image differences. It is able to automatically register remote sensing images under various conditions without human intervention. The main contributions are summarized as follows:

\begin{enumerate}
\item{An image registration algorithm named Multi-scale Histogram of Local Main Orientation (MS-HLMO) is designed to cope with various multi-source remote sensing images with multi-modal differences. Through comprehensive experimental verification, MS-HLMO is able to effectively deal with common multi-source remote sensing image registration, including multi-sensor, multi-temporal, and multi-viewpoint image pairs with different resolutions and data size, which reflects strong robustness and generalization.}

\item{A local feature named Histogram of Local Main Orientation (HLMO) is proposed, in which a basic feature map called Partial Main Orientation Map (PMOM) is developed for local feature extraction, which handles multi-modal differences and provides robust orientation information. In addition, HLMO utilizes a generalized GLOH-like (GGLOH) feature descriptor to extract local features. The proposed HLMO is invariant to intensity and rotation, and has outstanding ability in robust feature extraction from multi-source images.}

\item{To overcome the problem of scale differences in multi-source images, a multi-scale feature extraction and matching strategy based on Gaussian scale-space is applied in MS-HLMO, which has a good effect on solving scale differences and optimizing feature matching. MS-HLMO has a complete processing flow without manual intervention and is able to directly applied to the practical multi-source remote sensing joint analysis. To further research and assessment the proposed method, the code will be released on https://github.com/MrPingQi.}
\end{enumerate}

The rest of this paper is arranged as follows. Section II describes the overall framework of the proposed MS-HLMO and each critical process in detail. Section III presents experiments and discussion in terms of discrete and overall testing. Finally, this research is summarized in Section IV. 


\section{Proposed Registration Method}
\label{sec:method}
\begin{figure*}[!h]
 \centering
 \centerline{\includegraphics[width=17cm]{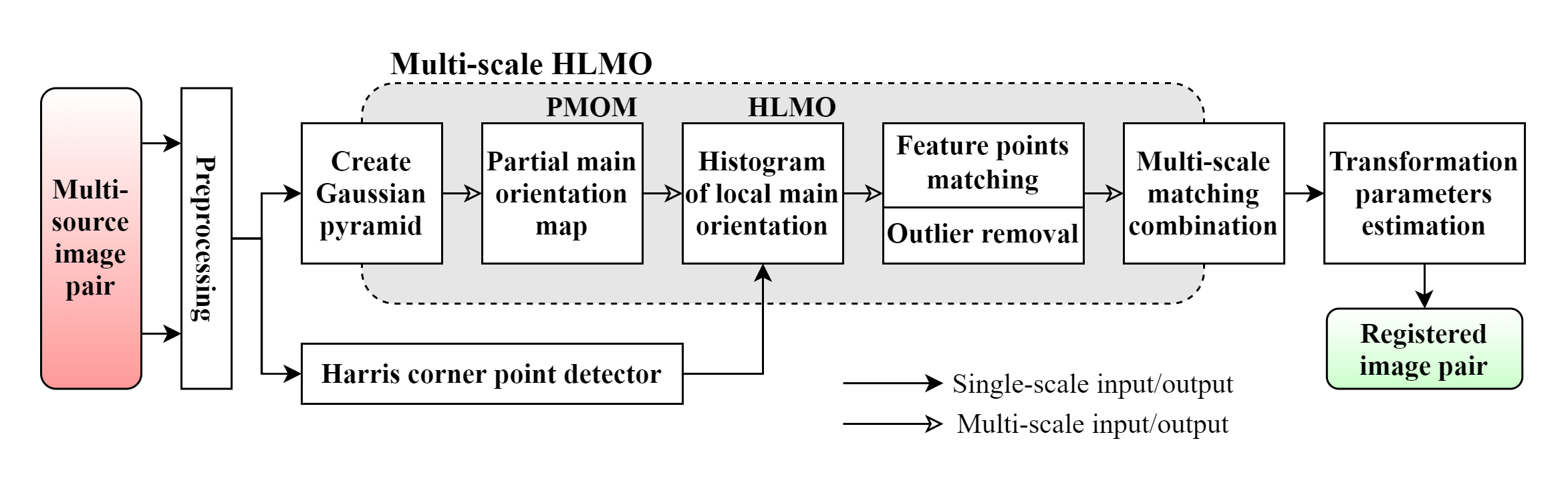}}
 \caption{The proposed multi-source remote sensing image registration framework of MS-HLMO, including Harris feature point detection, Histogram of Local Main Orientation feature extraction, and multi-scale registration strategy.}
 \label{fig:framework}
\end{figure*}

The framework of the proposed MS-HLMO registration algorithm is shown in Fig.\ref{fig:framework}. The input multi-source image pair to be registered is preprocessed, which includes data normalization and basic denoising. Then, the preprocessed single-band images are used for feature points detection and feature extraction. Harris corner point detection, which contains detail treatments for multi-source images, is adopted to generate feature points between the image pair for matching. The key process of the proposed HLMO feature extraction is carried out in a multi-scale strategy, in which Gaussian pyramids are built to create a scale-space of the images. The HLMO feature descriptors of each Harris corner point are extracted on the PMOM of the images. The feature points between the image pair are then matched according to the descriptors, and Fast Sample Consensus (FSC) is carried out to remove the outliers. The matching results in the scale-space are combined through a multi-scale matching strategy. Finally, the spatial transformation between the original image pair is determined by the coordinate relationship between matched feature points according to a selected transformation model.

\subsection{Harris Feature Point Detection}
\label{ssec:subhead}
Harris corner \cite{harris1988combined} is one of the most stable feature points, which is slightly affected by intensity and scale difference and has high computational efficiency \cite{gao2021multi,2021Multi}. It has the advantage in multi-source remote sensing images with multi-modal properties and large data size. Here, the similar strategy \cite{2021Multi} is used for feature points detection. The Harris corner response of each pixel is calculated by:
\begin{equation}
cornerness = \frac{{{\rm{det}}(\textbf{\emph{M}})}}{{{\rm{tr}}(\textbf{\emph{M}})}}
\end{equation}
\begin{equation}
\textbf{\emph{M}} = \left[ {\begin{array}{*{20}{l}}
{\sum\limits_{{\bf{W}_\sigma}} {{{\bf{G}}_x}^2} }&{\sum\limits_{{\bf{W}_\sigma}} {{{\bf{G}}_x} {{\bf{G}}_y}}}\\
{\sum\limits_{{\bf{W}_\sigma}} {{{\bf{G}}_x} {{\bf{G}}_y}} }&{\sum\limits_{{\bf{W}_\sigma}} {{{\bf{G}}_y}^2}}
\end{array}} \right]
\end{equation}
where $\rm{det}(\textbf{\emph{M}})$ and $\rm{tr}(\textbf{\emph{M}})$ are the determinant and trace of $\textbf{\emph{M}}$, respectively, ${{\bf{G}}_x}$ and ${{\bf{G}}_y}$ are the image's gradient along $x$ and $y$ directions, respectively, and $\bf{W}_\sigma$ is a Gaussian window with variance $\sigma$. Pixels with strong response are considered to be feature points with distinct structure and stability between multi-source images.

An important issue in practical multi-source remote sensing image registration is that the data size and scale relations between images to be registered are diverse. For example, an image with high resolution covers a smaller spatial area. Many of the existing algorithms only deal with the ideal case that the image pair has the same scale and size. This paper focuses on solving several key problems at the same time, that is, two uncertain factors of image scale and size should be considered simultaneously. The proposed MS-HLMO adopts local non-maximum suppression (LNMS) to solve this problem. Since the size and scale difference of the image pairs are uncertain, in Harris corner detection, it is expected that the feature points in the image pair are distributed as uniformly as possible with the use of LNSM. Then, the ratio of the window size in LNMS is set depending on the ratio of the data size of the image pair:
\begin{equation}
ratio = \sqrt {\frac{{M \times N}}{{m \times {\rm{n}}}}}
\end{equation}
where $M,N$ and $m,n$ are the length and width of the two images, respectively.

\begin{figure}[!h]
    \centering
        \subfloat[]{\includegraphics[width=2.5in]{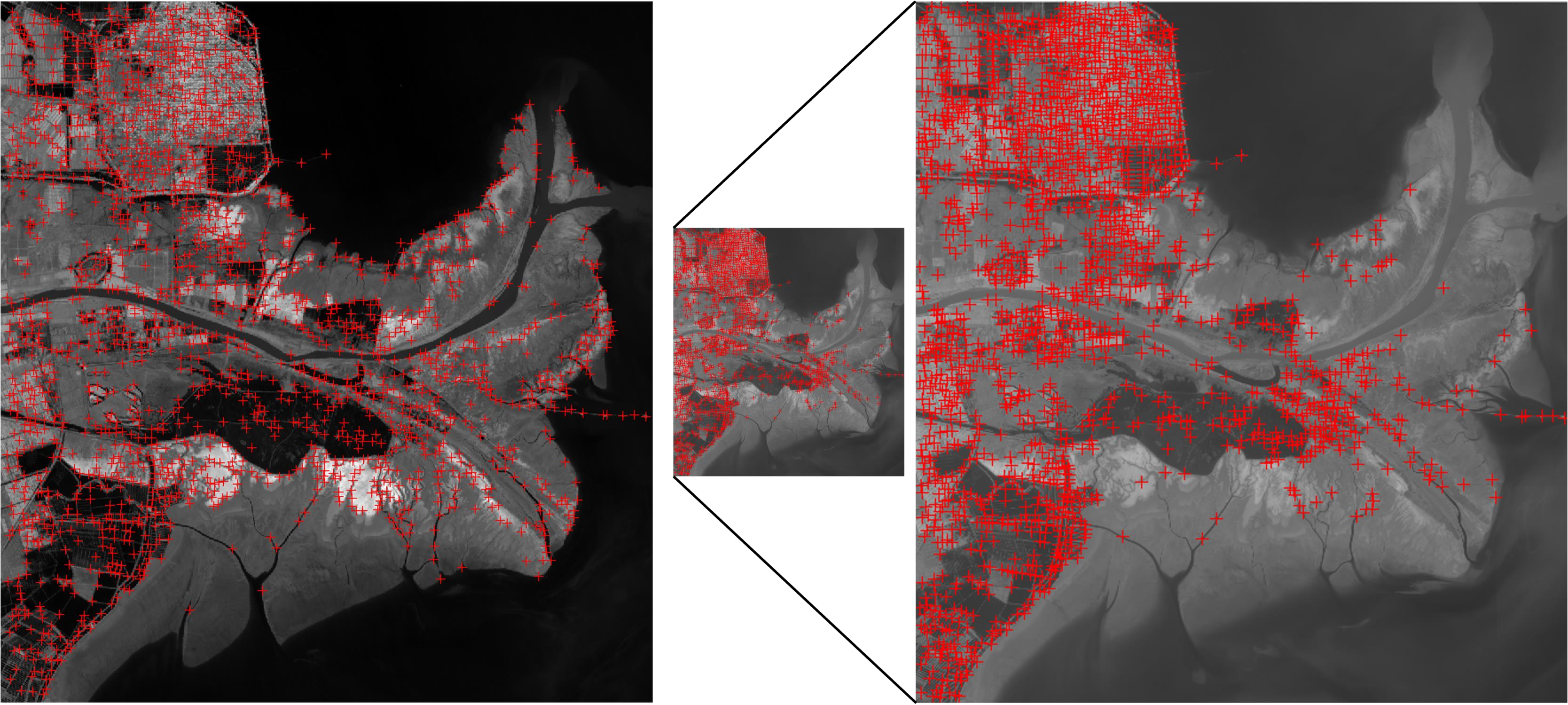}
        \label{fig:harris:a}}
    \hfil
        \subfloat[]{\includegraphics[width=2.5in]{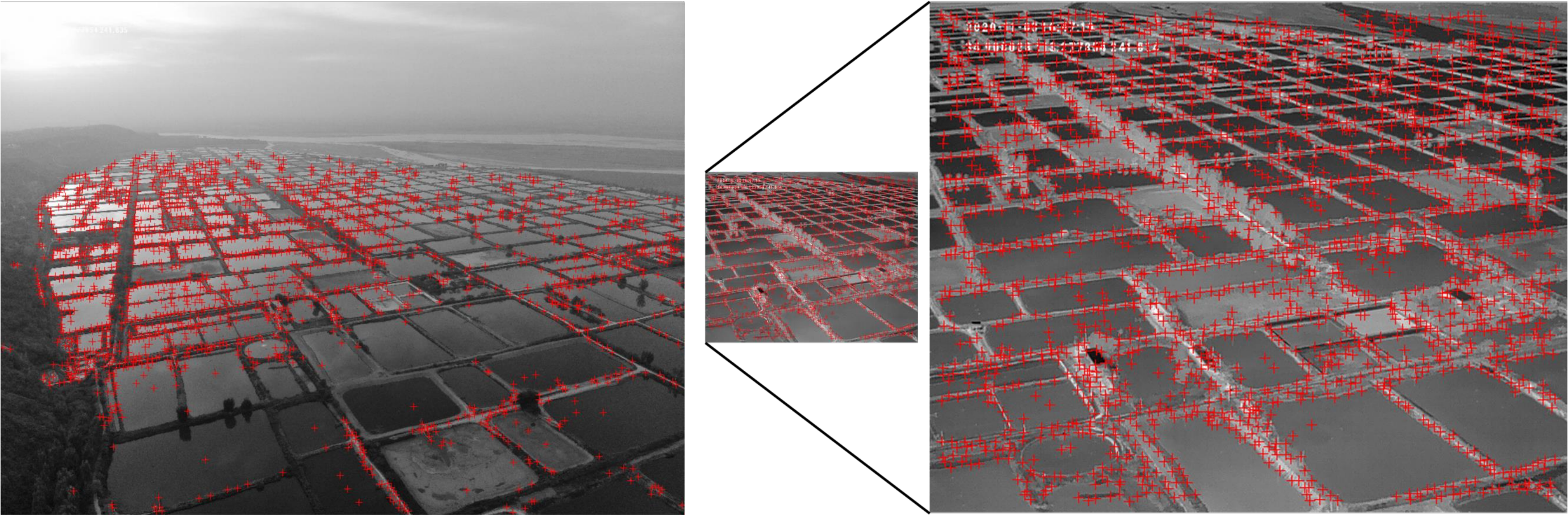}
        \label{fig:harris:b}}
    \caption{Examples of feature points detection results with LNMS. (a) Detection results of the image pair with different scale. (b) Detection results of the image pair with different scale and size.}
    \label{fig:harris}
\end{figure}

Consequently, the distribution of feature points is consistent with the images' scale proportion when there are scale differences, as shown in Fig.\ref{fig:harris:a}. When there is a size difference, the feature points are far more uniformly distributed, and the repeatability is higher, as shown in Fig.\ref{fig:harris:b}.

\subsection{Histogram of Local Main Orientation}
\label{ssec:subhead}

\subsubsection{Partial main orientation map}
Feature-point-based registration algorithms often use a local descriptor to extract the neighborhood information of the keypoints, and generate their feature vectors for similarity matching. For example, SIFT \cite{lowe2004distinctive}, HOG \cite{dalal2005histograms}, SURF \cite{bay2006surf} and PIIFD \cite{chen2010partial} employ gradient information as the basic feature. However, the performance of these algorithms is greatly degraded when processing multi-source, especially multi-sensor images. So, it is critical to extract invariant feature that is robust to ${{\cal F}_{\rm{Int}}}( \bullet )$, ${{\cal F}_{\rm{Rot}}}( \bullet )$, and ${{\cal F}_{\rm{Chg}}}( \bullet )$ for feature points description. A partial main orientation map (PMOM) is designed as the feature map in MS-HLMO, in which the Average Squared Gradient (ASG)\cite{kass1987analyzing} is adopted.

The ASG is a gradient weighting method. The elementary gradient of the image along $x$ and $y$ directions, i.e., ${\bf{G}}_x$ and ${\bf{G}}_y$ are calculated as
\begin{equation} \label{eqPMOM1}
\left[ \begin{array}{l}
{{\bf{G}}_x}(x,y)\\
{{\bf{G}}_y}(x,y)
\end{array} \right] = \left[ \begin{array}{l}
\frac{\partial }{{\partial x}}{\bf{I}}(x,y)\\
\frac{\partial }{{\partial y}}{\bf{I}}(x,y)
\end{array} \right]
\end{equation}
where ${\bf{I}}(x,y)$ represents the single-layer gray-scale image. The magnitude and orientation of its gradient, i.e., ${\bf{G}}_\rho$ and ${\bf{G}}_\varphi$ are
\begin{equation} \label{eqPMOM2}
\left[ \begin{array}{l}
{{\bf{G}}_\rho}\\
{{\bf{G}}_\varphi}
\end{array} \right] = \left[ \begin{array}{l}
\sqrt {{{\bf{G}}_x}^2 + {{\bf{G}}_y}^2} \\
\arctan \frac{{\bf{G}}_y}{{\bf{G}}_x}
\end{array} \right]\end
{equation}

In the ASG, a locally weighted squared gradient \cite{kass1987analyzing} along $x$ and $y$ directions, ${\bf{G}}_{{{\bf{W}}_\sigma},s,x}$ and ${\bf{G}}_{{\bf{W}_\sigma},s,y}$ are obtained as
\begin{equation} \label{eqPMOM3}
\left[ \begin{array}{l}
{{\bf{G}}_{{{\bf{W}}_\sigma},s,x}}\\
{{\bf{G}}_{{{\bf{W}}_\sigma},s,y}}
\end{array} \right] = \left[ \begin{array}{l}
\sum\limits_{{\bf{W}}_\sigma} {{{\bf{G}}_x}^2 - {{\bf{G}}_y}^2} \\
\sum\limits_{{\bf{W}}_\sigma} {2{{\bf{G}}_x}{{\bf{G}}_y}}
\end{array} \right]
\end{equation}
where ${\bf{W}}_\sigma$ is a Gaussian window with variance $\sigma$. Accordingly, the orientation of this gradient is
\begin{equation} \label{eqPMOM4}
{{\bf{G}}_{{{\bf{W}}_\sigma },s,\varphi }} = \angle ({{\bf{G}}_{{{\bf{W}}_\sigma },s,x}},{{\bf{G}}_{{{\bf{W}}_\sigma },s,y}})
\end{equation}
where $\angle (X,Y)$ is defined as
\begin{equation} \label{eqPMOM5}
\angle (X,Y) = \left\{ \begin{array}{l}
\arctan (\frac{Y}{X}), X \ge 0\\
\arctan (\frac{Y}{X}) + \pi , X < 0,Y \ge 0\\
\arctan (\frac{Y}{X}) - \pi , X < 0,Y < 0
\end{array} \right.
\end{equation}
making ${\bf{G}}_{{{\bf{W}}_\sigma },s,\varphi }$ within $(-\pi,\pi)$. According to \cite{kass1987analyzing}, this gradient is obtained by doubling the angle of the original gradient, so the orientation of the ASG is
\begin{equation} \label{eqPMOM6}
{{\bf{G}}_{{{\bf{W}}_\sigma},\varphi }} = \frac{1}{2}{{\bf{G}}_{{{\bf{W}}_\sigma},s,\varphi }}
\end{equation}

Compared with the classical gradient operator, the ASG orientation ${{\bf{G}}_{{{\bf{W}}_\sigma},\varphi}}\in(-\frac{\pi}{2},\frac{\pi}{2})$ reflects the weighted gradient orientation of a local region ${\bf{W}}_\sigma$, which is more robust and stable. In addition, the $x$ direction gradient is constant, and this characteristic meets the requirement that not affected by the reversal of gradient in intensity difference. Note that when $\sigma$ increases, the scale of ASG increases, which makes the local orientation more invariant to intensity difference and noise, but the uniqueness of local features decreases. From this, the following function is defined:
\begin{equation} \label{eqPMOM7}
{{\bf{G}}_{PMOM}} = \frac{1}{2}\angle (\sum\limits_\sigma  {\sum\limits_{{\bf{W}}_\sigma } {{{\bf{G}}_x}^2 - {{\bf{G}}_y}^2}} ,\sum\limits_\sigma  {\sum\limits_{{\bf{W}}_\sigma } {2{{\bf{G}}_x}{{\bf{G}}_y}} } )
\end{equation}
where a series of scale $\sigma$ are preset, the weighted responses ${\bf{G}}_{{{\bf{W}}_\sigma},s,x}$ and ${\bf{G}}_{{{\bf{W}}_\sigma},s,y}$ at each scale are added, and the ASG orientation is obtained. By filtering the image with Eq.(\ref{eqPMOM7}), the PMOM is obtained, where its value reflects the overall orientation of multiple scales in each partial area of the image.

\begin{figure}[h!]
 \begin{center}
  \includegraphics[width=3.5in]{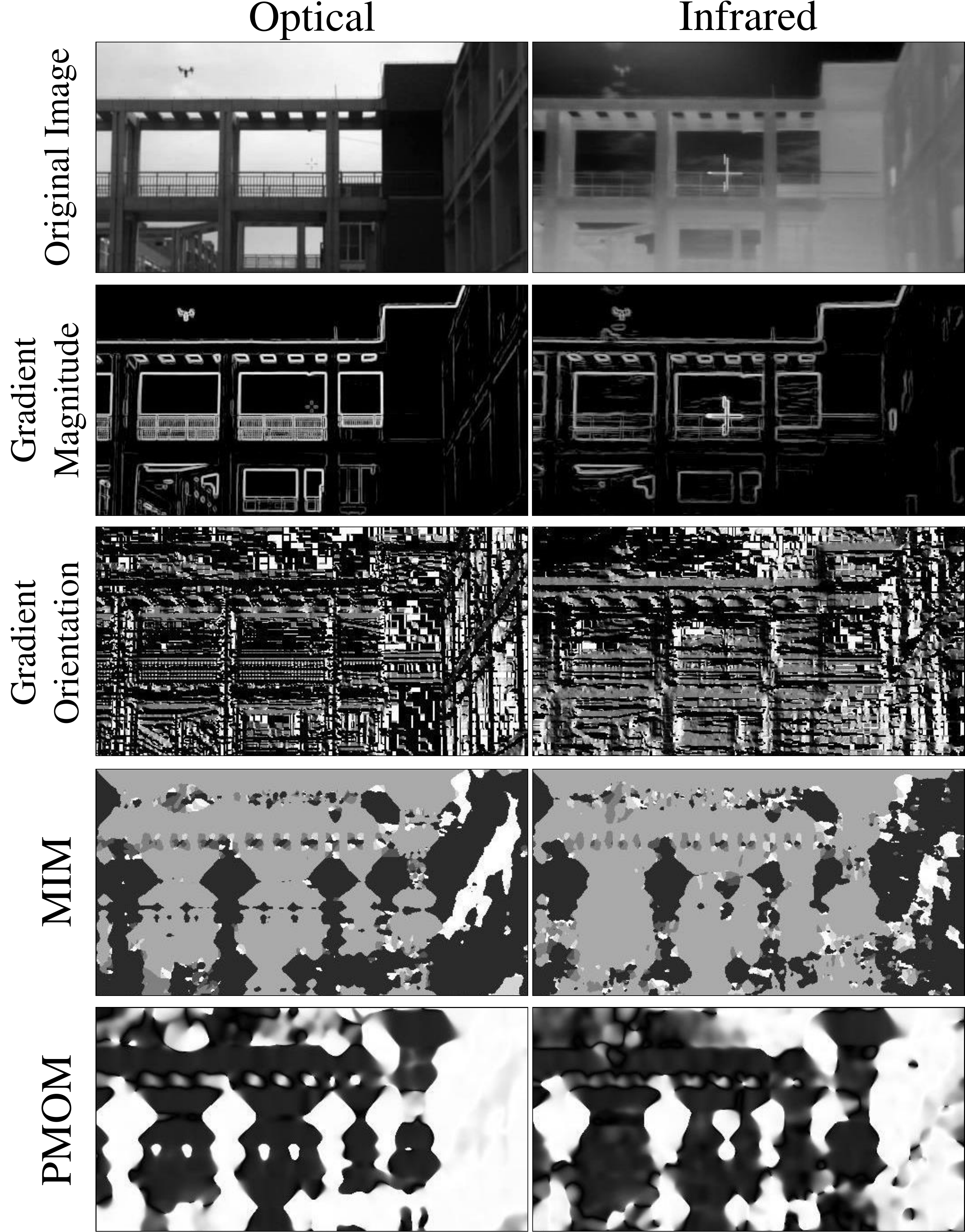}
  \caption{Comparison of feature maps of a selected scene, including a visible and an infrared image.}
  \label{fig:maps}
 \end{center}
\end{figure}

A visualized comparison of PMOM with other feature maps of typical multi-source data is shown in Fig.\ref{fig:maps}. The original data is a visible-infrared image pair, which contains obvious intensity difference. For comparison purposes, the images have been registered manually, basically eliminating the spatial differences of scale, rotation, and size. The magnitude and orientation of images' gradient are shown in Fig.\ref{fig:maps}, which are obtained using Eqs.(\ref{eqPMOM1})-(\ref{eqPMOM2}). These are the basic feature information in most algorithms \cite{lowe2004distinctive,dalal2005histograms,bay2006surf,chen2010partial,2021Multi}. It is observed that, due to the multi-modal properties of the original image pair, these two feature maps have large differences and instability, which is the main reason for the failure of most algorithms. The MIM \cite{li2019rift} of RIFT shown in Fig.\ref{fig:maps} also focuses on the local orientation of the image, where the maximum index is the main orientation among several ones. Compared with the directional gradient, Gabor transformation has a more stable response, which leads to RIFT robust to intensity difference. However, its value will also mutate due to local changes in images, and the rotation invariance is slightly poor. The image pair's PMOMs are shown at the bottom of Fig.\ref{fig:maps}. Compared with MIM, the proposed PMOM is not only more robust and stable between multi-modal images, but also continuous in value, which is conducive to achieving effective rotation invariance. In HLMO, PMOM is used as the unique feature information to extract local features of keypoints that are invariant to multi-modal properties.

\subsubsection{Descriptor extraction}

\begin{figure}[!h]
    \centering
        \subfloat[]{\includegraphics[width=2.2in]{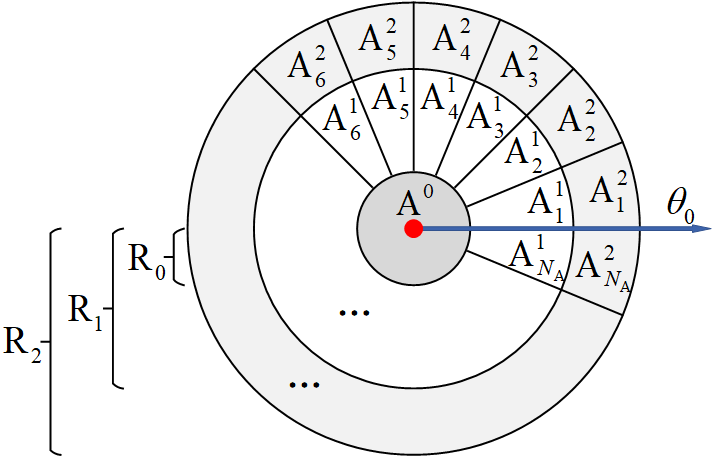}%
        \label{fig:des:a}}
    \hfil
        \subfloat[]{\includegraphics[width=1.1in]{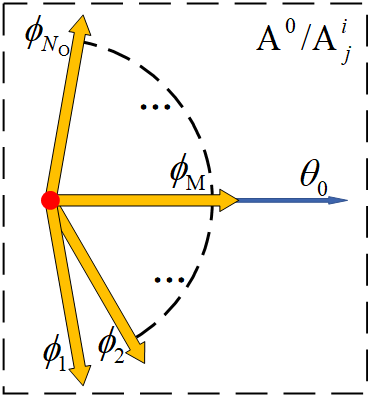}%
        \label{fig:des:b}}
    \caption{Descriptor structure of the proposed GGLOH. (a) Subregion partition of the local neighborhood of a feature point. (b) Angle quantification within each subregion.}
    \label{fig:des}
\end{figure}

\begin{figure}[h!]
 \begin{center}
  \includegraphics[width=3.3in]{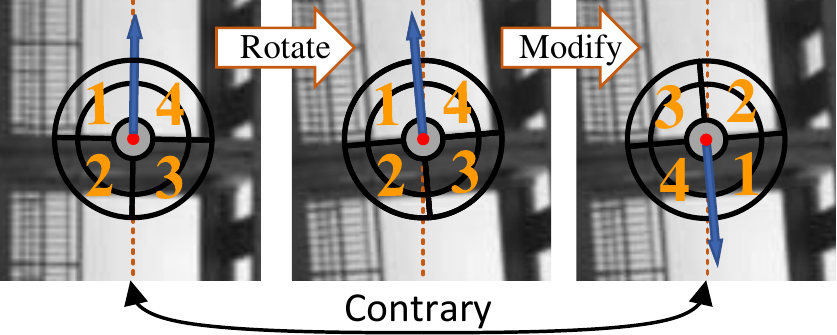}
  \caption{The problem caused by the jump of main orientation near $-\frac{\pi}{2}$ or $\frac{\pi}{2}$.}
  \label{fig:des180}
 \end{center}
\end{figure}

After determining the feature points and the specific feature for discrimination, the next step is to make use of the local feature information around each point and generate descriptors. Gradient Location and Orientation Histogram (GLOH) has shown excellent ability through experiments \cite{mikolajczyk2005performance}, and has been applied in multi-source remote sensing image registration \cite{dellinger2014sar,ma2016remote}. The original GLOH descriptor is a circular region divided by three circles, similar to that shown in Fig.\ref{fig:des180}, in which the two outer circular regions are divided into 4 parts, and the radius of the circular region divided are 5, 9, and 11. The partition size and the number are then improved \cite{mikolajczyk2005performance,dellinger2014sar,ma2016remote}. However, different parameters have various effects when treating multifarious types of images. In addition, if the number of outer ring regions is too small, the character of feature points is not significant, which makes it difficult to match accurately. If it is too large, the features are unstable, and the dimension of the descriptor is too high, which increases the burden of redundant calculation. To deal with this, a generalized GLOH-like (GGLOH) descriptor is proposed, and its structure of which is shown in Fig.\ref{fig:des:a}.

Let ${{\rm{A}}^0}$ denote the central circular region, and ${\rm{A}}_j^i,(i=1,2, j=1,...,{N_{\rm{A}}})$ represent the sector subregion $j$ in the outer ring region $i$. Let ${N_{\rm{A}}}$ be the number of the subregions in each out ring region, which is even, ${\theta _{\rm{0}}}$ be the main orientation of the feature point, and $R_0$, $R_1$, $R_2$ be the radii of the central and outer regions, respectively. Note that the orientations of pixels’ gradient in each region are counted as feature information, therefore, fair use of information in each region is expected. The number of pixels in each region should be roughly the same, and the weight of the outer regions should not change due to the change of ${N_{\rm{A}}}$. So the area of each region should be the same, that is
\begin{equation} \label{eqggloh}
{N_{\rm{A}}} \cdot \pi {\rm{R}}_{\rm{0}}^2 = \pi ({\rm{R}}_{\rm{1}}^2 - {\rm{R}}_{\rm{0}}^2) = \pi ({\rm{R}}_{\rm{2}}^2 - {\rm{R}}_{\rm{1}}^2)
\end{equation}
which also fixes the relationship between $R_0$, $R_1$, $R_2$ and ${N_{\rm{A}}}$. When ${N_{\rm{A}}}$ is given different values, the stability and importance of each region’s feature remains the same. In HLMO, the GGLOH is used to extract local features on the PMOM, where the orientation values within $(-\frac{\pi}{2},\frac{\pi}{2})$ are uniformly quantified to ${N_{\rm{O}}}$ values, as shown in Fig.\ref{fig:des:b}, where ${\phi_k}(k=1,2,...,{N_{\rm{O}}})$ is the angle after quantization. A histogram with ${N_{\rm{O}}}$ bins is obtained in each region.

It is simple to achieve rotation invariance of HLMO. For each keypoint, the PMOM value at its location is the main orientation, that is, the reference orientation ${\theta _0}$ of the GGLOH. Then, all of the PMOM values within the local area of GGLOH also take ${\theta _0}$ as the reference (0°), that is, all angle values minus ${\theta _0}$, and those beyond $(-\frac{\pi}{2},\frac{\pi}{2})$ are flipped to their opposite angles.

Another key problem is that the rotation and nonlinear intensity difference may cause the jump of the main orientations of some feature points near $-\frac{\pi}{2}$ and $\frac{\pi}{2}$. An example is shown in Fig.\ref{fig:des180}. In PIIFD \cite{chen2010partial}, a similar problem has been discovered and improvement has been made for SIFT. Then a similar strategy is adopted to process GLOH-like descriptors,

\begin{equation}
{{\bf{D}}_1} = \left[ {\begin{array}{*{20}{c}}
{\begin{array}{*{20}{c}}
{{\bf{H}}_1^1}&{{\bf{H}}_2^1}& \cdots &{{\bf{H}}_{{{{N_{\rm{A}}}} \mathord{\left/
 {\vphantom {{{N_{\rm{A}}}} 2}} \right.
 \kern-\nulldelimiterspace} 2}}^1}
\end{array}}\\
{\begin{array}{*{20}{c}}
{{\bf{H}}_1^2}&{{\bf{H}}_2^2}& \cdots &{{\bf{H}}_{{{{N_{\rm{A}}}} \mathord{\left/
 {\vphantom {{{N_{\rm{A}}}} 2}} \right.
 \kern-\nulldelimiterspace} 2}}^2}
\end{array}}
\end{array}} \right]
\end{equation}

\begin{equation}
{{\bf{D}}_2} = \left[ {\begin{array}{*{20}{c}}
{\begin{array}{*{20}{c}}
{{\bf{H}}_{{{{N_{\rm{A}}}} \mathord{\left/
 {\vphantom {{{N_{\rm{A}}}} 2}} \right.
 \kern-\nulldelimiterspace} 2} + 1}^1}&{{\bf{H}}_{{{{N_{\rm{A}}}} \mathord{\left/
 {\vphantom {{{N_{\rm{A}}}} 2}} \right.
 \kern-\nulldelimiterspace} 2} + 2}^1}& \cdots &{{\bf{H}}_{{N_{\rm{A}}}}^1}
\end{array}}\\
{\begin{array}{*{20}{c}}
{{\bf{H}}_{{{{N_{\rm{A}}}} \mathord{\left/
 {\vphantom {{{N_{\rm{A}}}} 2}} \right.
 \kern-\nulldelimiterspace} 2} + 1}^2}&{{\bf{H}}_{{{{N_{\rm{A}}}} \mathord{\left/
 {\vphantom {{{N_{\rm{A}}}} 2}} \right.
 \kern-\nulldelimiterspace} 2} + 2}^2}& \cdots &{{\bf{H}}_{{N_{\rm{A}}}}^2}
\end{array}}
\end{array}} \right]
\end{equation}

\begin{equation} \label{eqdes}
{\bf{D}} = \left[ {\begin{array}{*{20}{c}}
{{{\bf{D}}_1} + {{\bf{D}}_2}}\\
{c\left| {{{\bf{D}}_1} - {{\bf{D}}_2}} \right|}
\end{array}} \right]
\end{equation}
where ${\bf{H}}_j^i$ is the histogram vector of gradient orientation of region ${\rm{A}}_j^i$. In this way, no matter whether the main orientation of feature points is reversed 180° or not, descriptor ${\bf{D}}$ is composed of the addition and subtraction of the upper and lower parts of GGLOH according to the main orientation axis, without changing the regions' order. Finally, a descriptor vector ${\bf{D}}_P$ is generated for the feature point $P$, whose dimension is $(2 \cdot {N_{\rm{A}}}+1) \cdot {N_{\rm{O}}}$.


\subsection{Multi-scale Registration Strategy}
\label{ssec:subhead}
Scale difference ${{\cal F}_{{\rm{Sc}}}}( \bullet )$ of multi-source images has a great influence on local features. Some algorithms have quantitative scale judging methods, such as SIFT \cite{lowe2004distinctive} and LHOPC. However, it is found that these methods are invalid in images with large modal differences. The reason is that when images do not belong to the same degradation model, it is not credible to judge the scale quantitatively through local image feature information. Multi-source images often have scale differences, and sometimes the scale proportion is unknown. In order to deal with this key problem and realize scale robustness, a multi-scale feature extraction and matching strategy is designed in MS-HLMO.

\begin{figure}[h!]
 \begin{center}
  \includegraphics[width=2.5in]{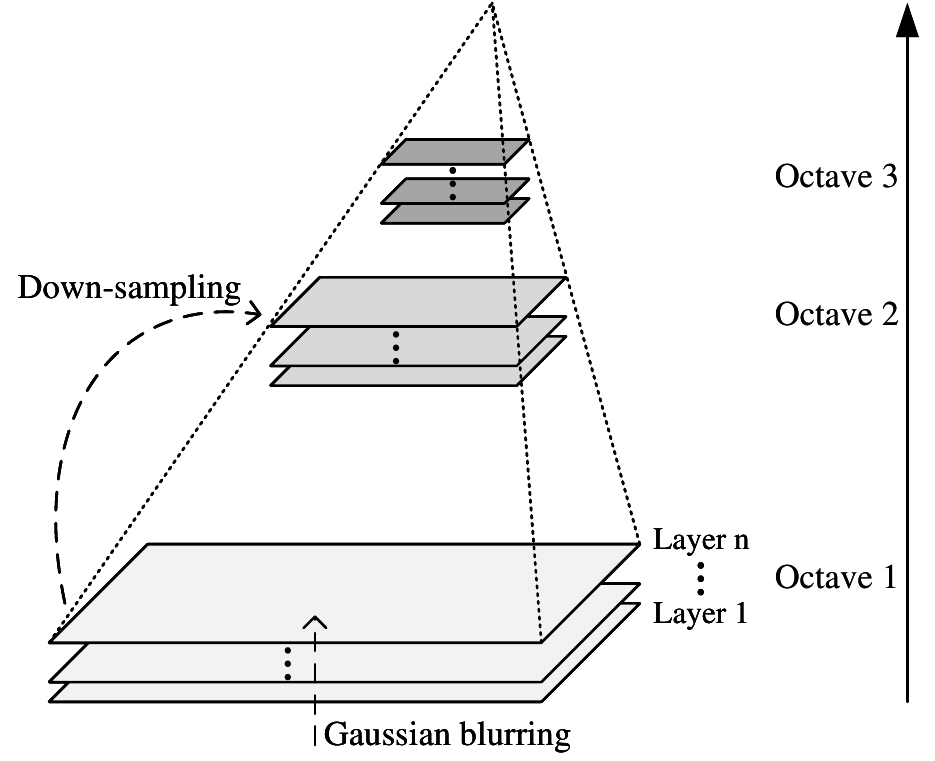}
  \caption{Structure of the Gaussian pyramid used in MS-HLMO.}
  \label{fig:pyramid}
 \end{center}
\end{figure}

Local information of feature points is extracted in the scale-space of the images. Based on the scale-space theory \cite{lindeberg1994scale}, the method of building image's Gaussian pyramids is adopted. The schematic diagram of establishing Gaussian pyramid of the image in the proposed algorithm is shown in Fig.\ref{fig:pyramid}. The original image is first sampled down step by step to obtain a series of images with different resolutions, that is, the first layer in each octave. Then in each octave, a series of Gaussian blurs are performed:
\begin{equation}\label{eqGauss1}
{\bf{L}} = {\bf{G}} * {\bf{I}}
\end{equation}
\begin{equation}\label{eqGauss2}
{\bf{G}} = \frac{1}{{\sqrt {2\pi {\sigma ^2}} }}{e^{\frac{{-({x^2} + {y^2})}}{{2{\sigma ^2}}}}}
\end{equation}
where ${\bf{I}}$ is the original image, ${\bf{G}}$ is a Gaussian kernel with a standard deviation of $\sigma$, and ${\bf{L}}$ is the Gaussian blur image.

\begin{algorithm}[htp]
\caption{\label{multi1} \footnotesize Proposed MS-HLMO Feature Extraction}
\begin{algorithmic}
\footnotesize
\STATE {\bf Input:} single-band image $\mathbf{I}$, feature point set $P_{\mathbf{I}}$, total number of octaves ${N_{\rm{GO}}}$ and layers ${N_{\rm{GL}}}$ in Gaussian pyramid, subregion and angle partition parameters ${N_{\rm{A}}}$, ${N_{\rm{O}}}$ in GGLOH , patch size $S$ of HLMO.
\STATE Through down-sampling and Eq.(\ref{eqGauss1})(\ref{eqGauss2}), the Gaussian pyramid ${\bf{G}}_{\bf{I}}(O,L)$ of image $\mathbf{I}$ is built with ${N_{\rm{GO}}}$ octaves and ${N_{\rm{GL}}}$ layers in each octave.
\STATE In each layer of ${\bf{G}}_{\bf{I}}(O,L)$:
\STATE \hspace*{0.1in}Calculate the PMOM of this layer to get ${\bf{F}}_{\bf{I}}(O,L)$ according to Eq.(\ref{eqPMOM1})(\ref{eqPMOM7})
\STATE \hspace*{0.1in}For each feature point $P$ in $P_{\bf{I}}$:
\STATE \hspace*{0.2in}Calculate the corresponding position
\STATE \hspace*{0.2in}Take the PMOM value at the position as the main orientation ${\theta_0}$
\STATE \hspace*{0.2in}Taking the main orientation as the reference direction ($0^{\circ}$), establish a GGLOH window with size (diameter) of S
\STATE \hspace*{0.2in}Statistics the PMOM value within each region of GGLOH to obtain the basic feature descriptor $D_{1}(P,O,L)$ and $D_{2}(P,O,L)$
\STATE \hspace*{0.2in}Obtain the descriptor $D(P,O,L)$ of $P$ with Eq.\ref{eqdes}.
\STATE {\bf Output:} feature descriptor set $D_{P_{\bf{I}}}(O,L)$
\end{algorithmic}
\end{algorithm}

After the scale-space of the images is established, for each Harris corner point, the HLMO descriptor is calculated by obtaining the local information at the corresponding location of each feature point in the scale-space. The proposed multi-scale HLMO feature extraction method is provided in Algorithm 1, where $O$ is the octave number in the Gaussian pyramid, and $L$ is the layer number. The algorithm outputs the feature point descriptor set $D_{P_{\bf{I}}}(O,L)$, which contains $(2 \cdot {N_{\rm{A}}}+1) \cdot {N_{\rm{O}}}$-dimensional vectors for each feature point at each scale.

\begin{figure}[h!]
 \begin{center}
  \includegraphics[width=3.5in]{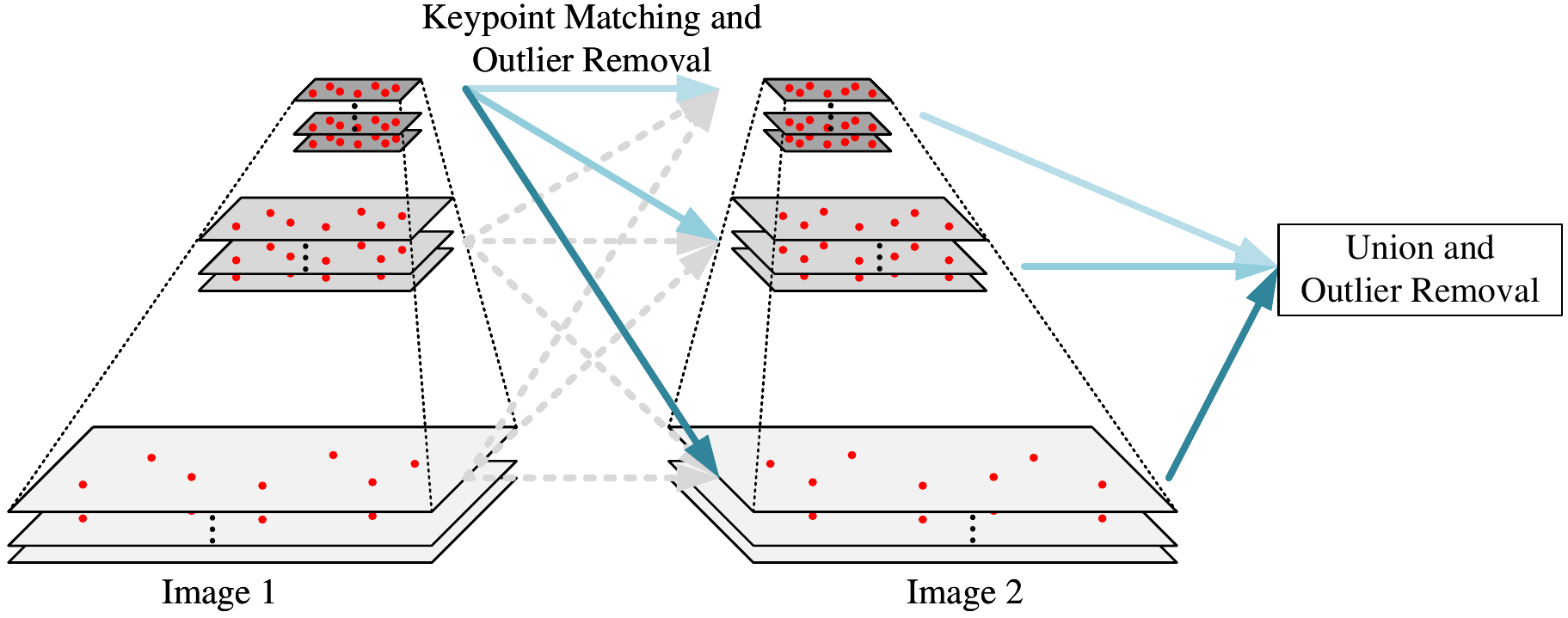}
  \caption{Multi-scale keypoints matching strategy in MS-HLMO.}
  \label{fig:pyramids}
 \end{center}
\end{figure}

The next step is to match the feature point sets of the image pair according to the descriptor sets. The process of the multi-scale feature matching is provided in Algorithm 2. In the process, each scale is matched in turn. Then the matching results are merged step by step while the outlier removal is carried out to realize the optimization of matching points. The final matching results are used to determine the spatial transformation between images. The most critical is to combine all the matching results of feature points and remove outliers, so as to maximize the correct matches of all scales. Fig.\ref{fig:pyramids} shows this process visually. Obviously, this is a general approach to handle all kinds of unknown scale differences. When the scale proportion of images is known or can be estimated, then the above process can be greatly simplified. In this case, the proposed multi-scale strategy still has the advantages of enhancing feature matching and maximizing the number of matching points.

\begin{algorithm}[htp]
\caption{\label{multi2} \footnotesize Proposed MS-HLMO Feature Matching}
\begin{algorithmic}
\footnotesize
\STATE {\bf Input:} feature point set of the image pair $P_{{\bf{I}}1}$, $P_{{\bf{I}}2}$, feature descriptor set of the image pair $D_{P_{{\bf{I}}1}}(O_{1},L_{1})$, $D_{P_{{\bf{I}}2}}(O_{2},L_{2})$
\STATE Take each layer of $D_{P_{{\bf{I}}1}}(O_{1},L_{1})$:
\STATE \hspace*{0.1in}Take each layer of $D_{P_{{\bf{I}}2}}(O_{2},L_{2})$:
\STATE \hspace*{0.2in}Match $P_{{\bf{I}}1}$ and $P_{{\bf{I}}2}$ using Euclidean distance of the descriptorss
\STATE \hspace*{0.2in}Remove outliers, producing the matching result of a single scale $M(O_{1},O_{2},L_{1},L_{2})$
\STATE The matching results of all layers in each octave of the scale-space are union and then optimized with outlier removal, producing the matching result $M_{L}(O_{1},O_{2})$
\STATE The matching results of all octaves in $M_{L}(O_{1},O_{2})$ are union and then optimized with outlier removal, producing the final matching result $M_{OL}(P_{{\bf{I}}1},P_{{\bf{I}}2})$
\STATE {\bf Output:} feature points matching set $M_{OL}(P_{{\bf{I}}1},P_{{\bf{I}}2})$
\end{algorithmic}
\end{algorithm}

\section{Experimental Results and Discussion}
\label{sec:results}
To validate the registration performance of the proposed framework, a set of typical multi-source remote sensing images are carefully selected, and six effective registration algorithms are used for comparison, including SIFT \cite{lowe2004distinctive}, SAR-SIFT \cite{ma2016remote}, PSO-SIFT \cite{ma2016remote}, PIIFD \cite{chen2010partial}, MS-PIIFD \cite{2021Multi}, and RIFT \cite{li2019rift}, all of which contain processes dealing with multi-modal properties. The experiments are implemented using MATLAB2021a on a platform with AMD-Ryzen-7-5800X 3.80GHz CPU, 32GB RAM.

\subsection{Experimental Data and Preprocessing}
\label{ssec:data}
The data used in experiments are multi-source remote sensing images of 17 scenes, labeled a$\sim$q. These images include hyperspectral image (HSI), multispectral (MSI), synthetic aperture radar (SAR), visible images, infrared images, depth maps such as LiDAR-derived DSM (digital surface model), and even artificial maps. They contain day-night image pair and images from different years. The scenes cover spaceborne, airborne, and ground-based remote sensing data. There are various intensity and geometric distortions with the images. Table \ref{tbl:data} shows the types and sizes of each set of data, as well as whether they have significant intensity difference, rotation, or scale difference. The specific form of the scenes is shown in Fig.\ref{fig:matching} and \ref{fig:checker}. Except for artificial maps, these are all real data without manual manipulation.

\begin{table}[h!]
 \centering
 \setlength{\tabcolsep}{1.2mm}
 \caption{Multi-source remote sensing data of 17 scenes.}
 \label{tbl:data}
 \begin{tabular}{ccccccc}
  \toprule
   \multicolumn{2}{c}{\multirow{2}{*}{Scene}} & \multirow{2}{*}{Type} & \multirow{2}{*}{Size} & Intensity  & \multirow{2}{*}{Rotation} & Scale \\
                      &                       &                       &                       & Difference &                           & Difference \\
  \midrule
   \multirow{6}{*}{a} & \multirow{3}{*}{a1} &      MSI & 3555$\times$4026 & \multirow{3}{*}{$\checkmark$} &  & \multirow{3}{*}{$\checkmark$} \\
                      &                     &      HSI & 1185$\times$1342 &  &  &  \\
                      &                     &      SAR & 3555$\times$4026 &  &  &  \\
          \cline{2-7} & \multirow{3}{*}{a2} &      MSI & 1287$\times$2035 & \multirow{3}{*}{$\checkmark$} &  & \multirow{3}{*}{$\checkmark$} \\
                      &                     &      HSI &  422$\times$678  &  &  &  \\
                      &                     &      SAR & 1366$\times$887  &  &  &  \\ \cline{1-7}
   \multicolumn{2}{c}{\multirow{2}{*}{b}}   &      MSI & 3529$\times$1756 & \multirow{2}{*}{$\checkmark$} &  & \multirow{2}{*}{$\checkmark$} \\
                      &                     &      HSI & 1175$\times$585  &  &  &  \\ \cline{1-7}
   \multicolumn{2}{c}{\multirow{2}{*}{c}}   &      MSI & 2461$\times$4139 & \multirow{2}{*}{$\checkmark$} &  & \multirow{2}{*}{$\checkmark$} \\
                      &                     &      MSI & 2271$\times$3152 &  &  &  \\ \cline{1-7}
   \multicolumn{2}{c}{\multirow{2}{*}{d}}   &      PAN & 1000$\times$1000 & \multirow{2}{*}{$\checkmark$} &  & \multirow{2}{*}{$\checkmark$} \\
                      &                     &      HSI &  550$\times$450  &  &  &  \\ \cline{1-7}
   \multicolumn{2}{c}{\multirow{2}{*}{e}}   &  visible &  614$\times$614  & \multirow{2}{*}{$\checkmark$} & \multirow{2}{*}{$\checkmark$} &  \\
                      &                     & infrared &  611$\times$611  &  &  &  \\ \cline{1-7}
   \multicolumn{2}{c}{\multirow{2}{*}{f}}   &  visible &  617$\times$593  & \multirow{2}{*}{$\checkmark$} & \multirow{2}{*}{$\checkmark$} &  \\
                      &                     & infrared &  617$\times$593  &  &  &  \\ \cline{1-7}
   \multicolumn{2}{c}{\multirow{2}{*}{g}}   &  visible & 1920$\times$1080 & \multirow{2}{*}{$\checkmark$} &  & \multirow{2}{*}{$\checkmark$} \\
                      &                     & infrared &  667$\times$504  &  &  &  \\ \cline{1-7}
   \multicolumn{2}{c}{\multirow{2}{*}{h}}   &  visible & 4056$\times$3040 & \multirow{2}{*}{$\checkmark$} &  & \multirow{2}{*}{$\checkmark$} \\
                      &                     & infrared &  640$\times$512  &  &  &  \\ \cline{1-7}
   \multicolumn{2}{c}{\multirow{2}{*}{i}}   &  visible &  633$\times$460  & \multirow{2}{*}{$\checkmark$} &  &  \\
                      &                     & infrared &  670$\times$508  &  &  &  \\ \cline{1-7}
   \multicolumn{2}{c}{\multirow{2}{*}{j}}   &      HSI &  349$\times$1905 & \multirow{2}{*}{$\checkmark$} &  &  \\
                      &                     &    LiDAR &  349$\times$1905 &  &  &  \\ \cline{1-7}
   \multicolumn{2}{c}{\multirow{2}{*}{k}}   &      HSI &  166$\times$600  & \multirow{2}{*}{$\checkmark$} &  &  \\
                      &                     &    LiDAR &  166$\times$600  &  &  &  \\ \cline{1-7}
   \multicolumn{2}{c}{\multirow{2}{*}{l}}   &      HSI &  325$\times$220  & \multirow{2}{*}{$\checkmark$} &  &  \\
                      &                     &    LiDAR &  325$\times$220  &  &  &  \\ \cline{1-7}
   \multicolumn{2}{c}{\multirow{2}{*}{m}}   &  visible &  500$\times$500  & \multirow{2}{*}{$\checkmark$} &  &  \\
                      &                     &    depth &  500$\times$500  &  &  &  \\ \cline{1-7}
   \multicolumn{2}{c}{\multirow{2}{*}{n}}   &  visible &  500$\times$500  & \multirow{2}{*}{$\checkmark$} &  &  \\
                      &                     &      map &  500$\times$500  &  &  &  \\ \cline{1-7}
   \multicolumn{2}{c}{\multirow{2}{*}{o}}   &  visible &  500$\times$500  & \multirow{2}{*}{$\checkmark$} &  &  \\
                      &                     &  visible &  500$\times$500  &  &  &  \\ \cline{1-7}
   \multicolumn{2}{c}{\multirow{2}{*}{p}}   &      SAR &  600$\times$500  & \multirow{2}{*}{$\checkmark$} & \multirow{2}{*}{$\checkmark$} &  \\
                      &                     &      SAR &  600$\times$500  &  &  &  \\ \cline{1-7}
   \multicolumn{2}{c}{\multirow{2}{*}{q}}   &  visible &  300$\times$300  &  & \multirow{2}{*}{$\checkmark$} &  \\
                      &                     &  visible &  300$\times$300  &  &  &  \\
  \bottomrule
 \end{tabular}
\end{table}

As MS-HLMO and other methods only focus on spatial rather than waveband features, while some types of images have multiple wavebands, such as MSI, HSI, or LiDAR, single-band gray-scale images need to be obtained from the original multi-band ones. It is acceptable to use data dimension reduction, such as Principal Component Analysis (PCA), to obtain single-band images. However, in multi-source registration, the images may have multi-modal properties. Due to its enhancing pixel-level features, data dimension reduction could break the stable and unified spatial structures of each type of ground covers in the image, and increase the structural differences between images. Then, registration would be more difficult as ${{\cal F}_{{\rm{Chg}}}}(\bullet)$ is exacerbated, which is undesired. Note that HLMO is not subject to the changes in image's grayscale, so it is optional to select any one of the bands as the input, or just simply add up all the bands, which has little effect on the registration result.

\subsection{Experiment Settings}
\label{ssec:para}
The proposed MS-HLMO contains seven main parameters that greatly affect the registration, which are the number of feature points, the scale of PMOM, the $R_2$, ${N_{\rm{A}}}$, and ${N_{\rm{O}}}$ of GGLOH, and ${N_{\rm{GO}}}$ and ${N_{\rm{GL}}}$ of the Gaussian pyramid. In general, the lower the number of subregions ${N_{\rm{A}}}$ and angles ${N_{\rm{O}}}$ are, the higher the robustness of features is, but the lower the separability of the feature points is. Conversely, the more distinct the features are, but the lower the stability is, which also increases the computational burden. In the experiment, both ${N_{\rm{A}}}$ and ${N_{\rm{O}}}$ are set to 12, and $R_2$ is set to 48, which has the best performance and stability in the majority of multi-source data. Then, $R_1$ and $R_0$ in GGLOH are determined by Eq.\ref{eqggloh}, and the patchsize of HLMO is $2\times R_2 = 96$. Through a simple set of tests, the threshold of the Harris corner points' number is set at 2000, and ${N_{\rm{GO}}}$ and ${N_{\rm{GL}}}$ are set to 3 and 4, which handles the registration effectively enough with high efficiency.

In the calculation of PMOM, 10 scales are set for fusion, of which the radiuses are evenly spaced between $R_0$ and $R_2$, and the $\sigma$ is 1/3 of each radius. In this way, when GGLOH is applied to a feature point, the main orientation of the patch, that is the PMOM value of the center point, is only determined by the pixels' values within the patch, excluding information beyond the GGLOH region. 10 scales are sufficient to fully obtain the local multi-scale weighted main orientation information in various multi-source images, which balances the robustness and uniqueness with a low computational burden.

The number of correct matches (NCM) is employed as the main evaluation metric of feature matching. The correct matches refer to the matched points that are basically at the same position in the real space, and the NCM is positively correlated with the accuracy of the transformation model. Note that if the NCM of an image pair is less than 3, the parameters of the spatial transformation model cannot be solved, then the registration process is considered to fail.

To ensure the fairness of the comparison, except for the feature point detection and descriptor extraction methods, all the processes are the same, including preprocessing, feature point matching, outlier removal, transformation, etc. The feature points used in PIIFD, MS-PIIFD, and RIFT are the same as the proposed algorithm with the strategy in \cite{2021Multi}. Different algorithms use various feature matching and outlier removal methods in their original process, which interferes with the comparison of registration effects. In this experiment, Euclidean distance measurement is used for all feature matching, and FSC is used for outlier removal for a fair comparison. As FSC has random property, the result fluctuates in the experiment, which is not very large but does have an impact. So each test is repeated 50 times and the result with the most NCM is taken.

\begin{figure*}[!h]
    \centering
        \subfloat[]{\label{fig:rot:a}
        \includegraphics[height=0.8in]{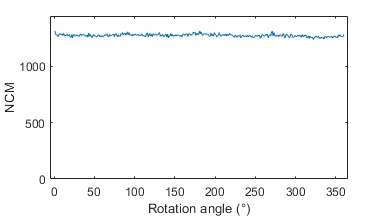}}
        \subfloat[]{\label{fig:rot:b}
        \includegraphics[height=0.8in]{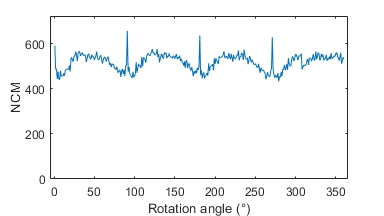}}
        \subfloat[]{\label{fig:rot:c}
        \includegraphics[height=0.8in]{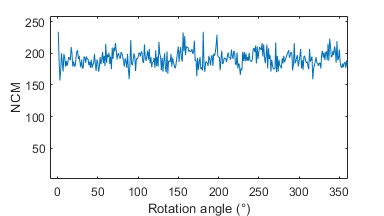}}
        \subfloat[]{\label{fig:rot:d}
        \includegraphics[height=0.8in]{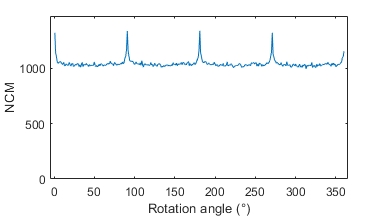}}
    \hfil
        \subfloat[]{\label{fig:rot:e}
        \includegraphics[height=0.8in]{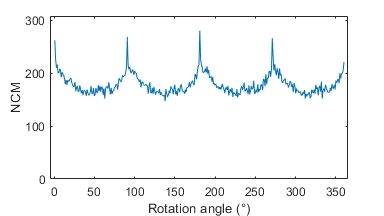}}
        \subfloat[]{\label{fig:rot:f}
        \includegraphics[height=0.8in]{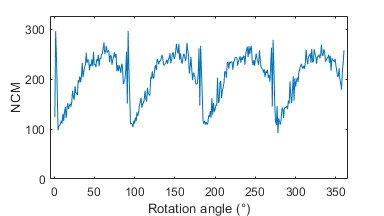}}
        \subfloat[]{\label{fig:rot:g}
        \includegraphics[height=0.8in]{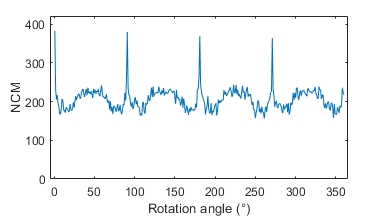}}
        \subfloat[]{\label{fig:rot:h}
        \includegraphics[height=0.8in]{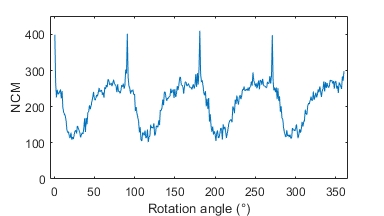}}
    \caption{NCMs of MS-HLMO on different types of multi-source remote sensing scenes as the rotation angles from 0° to 359°. (a) MSI-HSI. (b) MSI-SAR. (c) MSI-map. (d) PAN-HSI. (e) visible-infrared. (f) HSI-LiDAR. (g) visible-depth. (h) SAR-SAR.}
    \label{fig:rot}
\end{figure*}

\subsection{Invariance Tests}

The crux of multi-source image registration based on feature matching lies in various multimodal differences, and the robustness of local features to these differences is crucial. According to the analysis, intensity difference ${{\cal F}_{\rm{Int}}}( \bullet )$, rotation ${{\cal F}_{{\rm{Rot}}}}( \bullet )$, and scale difference ${{\cal F}_{{\rm{Sc}}}}( \bullet )$ are the most common problems. So we design experiments to singly test the invariance of MS-HLMO from these three aspects, so as to test whether it deals with each multimodal problem between multi-source images.

\subsubsection{Intensity invariance}
In order to independently test the intensity invariance of MS-HLMO, the influence of other factors is excluded. Several image pairs are selected and manually registered in advance, providing image pairs of the same size that basically eliminated the differences in scale and rotation as the input image of registration. The NCMs of the registration results are listed in Table \ref{tbl:intensity}.
The MS-HLMO$^+$ represents the proposed algorithm without the operation of rotation invariance, that is, the main orientation is not assigned to each feature point, and the reference orientation ${\theta _{\rm{0}}}$ in feature description is set to 0. In this way, the robustness to intensity difference is evaluated more accurately, and the influence of descriptor rotation and PMOM value changes is excluded. As RIFT is an effective feature focusing on intensity difference, it is chosen for comparison.

\begin{table}[h!]
 \centering
 \setlength{\tabcolsep}{1.5mm}
 \caption{NCMs of 8 scenes with only intensity difference by MS-HLMO and RIFT.}
 \label{tbl:intensity}
 \begin{tabular}{ccccccccc}
  \toprule
  \multirow{2}{*}{Method} & \multicolumn{8}{c}{Scene}\\
                            \cline{2-9} & a1 & a2 & d & j & g & m & a4 & p \\
  \midrule
   RIFT$^+$    & 1032 &  385 &  393 & 176 & 104 & 189 &  33 &  92 \\
   MS-HLMO     & 1417 &  620 & 1321 & 149 & 280 & 374 & 220 & 575 \\
   MS-HLMO$^+$ & 1434 & 1017 & 1454 & 771 & 351 & 846 & 386 & 868 \\
  \bottomrule
 \end{tabular}
\end{table}

The 8 scenes basically cover common remote sensing image types, including various optical images, SAR, depth map, and artificial map. From the numerical results, each group has obtained sufficient correct matches, especially in scene (a1) and (d), about 1400 of 2000 detected Harris corner points have been successfully matched. In most scenes, the NCMs of the MS-HLMO are far more than RIFT, reflecting stronger robustness to intensity difference. In scene (j), the MS-HLMO with rotation invariance obtained fewer matches than RIFT. Through analysis, this is because scene (j) is an HSI-LiDAR image pair with large size and complex content, where the modal difference between them is large. Thus, the stability of the feature points' main orientation is not high. The inconsistent reference direction leads to large differences in some feature points' descriptors, while RIFT does not have influence caused by the feature point's orientation. However, without rotation interference, the MS-HLMO$^+$ obtains a significant number of correct matches, more than four times to RIFT. The experiment shows that MS-HLMO with PMOM as its basic feature has strong intensity invariance and is very effective for multi-modal image registration.

\subsubsection{Rotation invariance}

\begin{figure}[!h]
 \begin{center}
  \includegraphics[width=3.5in]{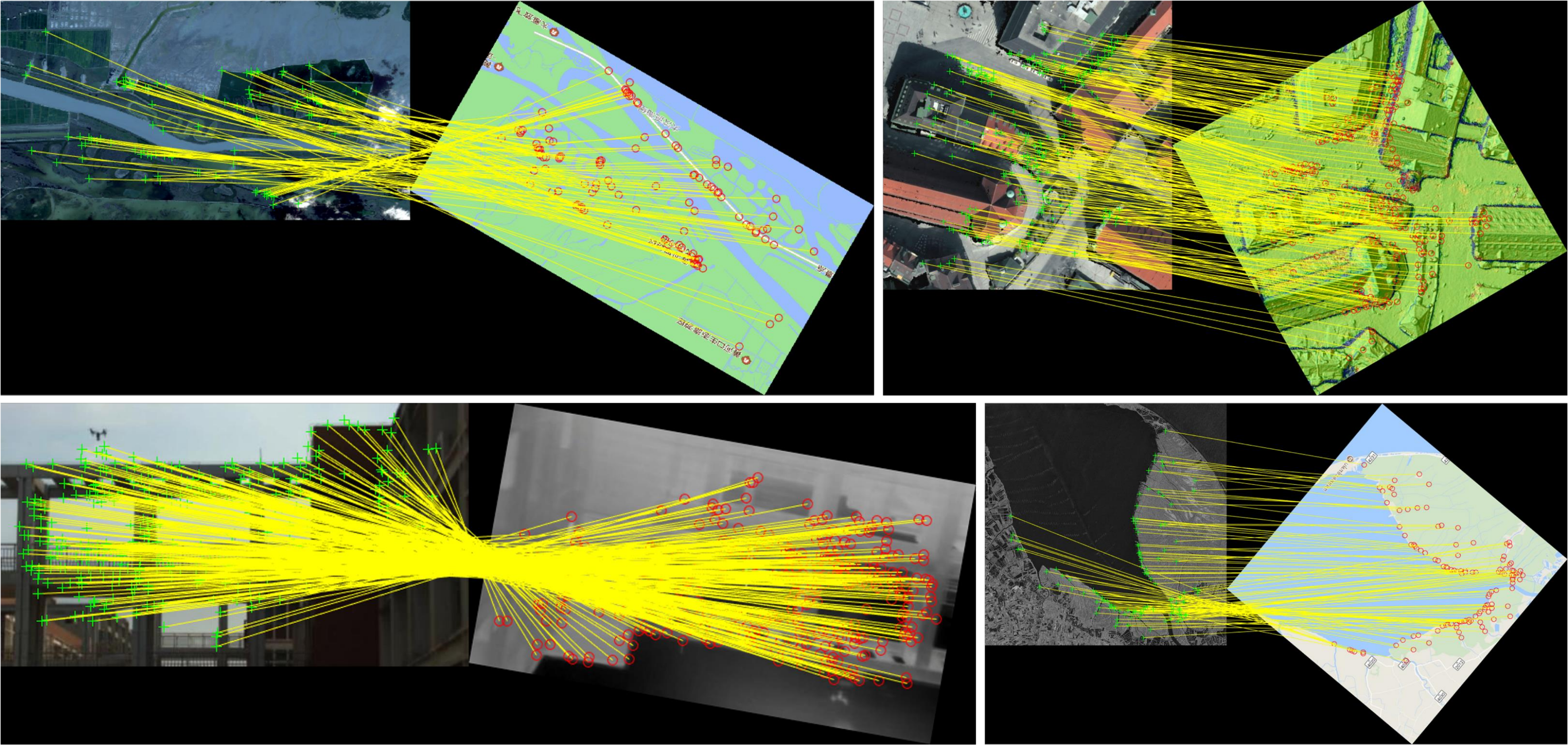}
  \caption{Examples of feature matching results with image rotation.}
  \label{fig:rot_r}
 \end{center}
\end{figure}

To test rotation invariance, for each one of the above image pairs without scale and rotation differences, one of them is fixed and the other is rotated. The rotation angles are from 0° to 359° with an interval of 1°, from which a total of 360 image pairs of each scene are obtained. Then, the registration algorithms are performed on each image pair, and the corresponding NCMs are plotted in Fig.\ref{fig:rot}. Samples of registration results under different rotation angles are shown in Fig.\ref{fig:rot_r}.

It is observed that the NCMs of the 8 scenes vary as the rotation angles change, but are all greater than 100. None of the image pairs fails to be registered at any angles, indicating that the proposed MS-HLMO has rotation invariance over the entire angle interval of 0° to 360°. In most scenes, as the images rotate, the NCMs fluctuate with a cycle of 90°, the reason of this phenomenon is that the image rotation involves data interpolation as a digital image is stored a matrix. In other words, the change of the local gradient orientation is not linear, which is inevitable in digital image processing. As a result, the main orientation of each feature point is not completely stable with a certain offset from the image rotation angle. Relatively, the rotation of 0°, 90°, 180°, and 270° do not change the data value, and the main orientation of each feature point will not shift, resulting in the four peaks in the curve. Theoretically, the four intervals evenly divided by 90° from 0° to 360° are completely consistent, that is, the descriptors of a feature point are completely the same when rotated by 90°, 180°, and 270° which is confirmed by the experimental results. MS-HLMO has a strong rotation invariance and fully copes with image rotation from various angles. In addition, the rotation invariance is not restricted by the types of imaging sensors.

\subsubsection{Scale invariance}
\begin{figure}[h!]
 \begin{center}
  \includegraphics[width=2.2in]{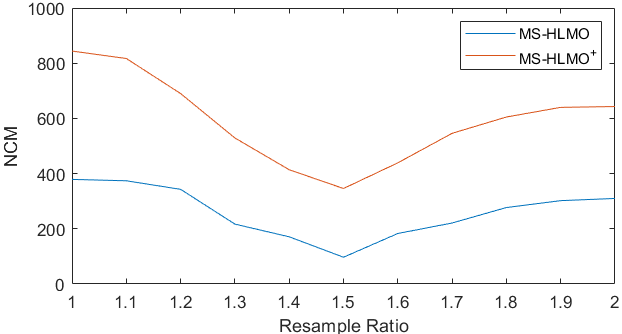}
  \caption{NCMs of MS-HLMO and MS-HLMO$^+$ on scene (m) as the scaling ratios from 1 to 2.}
  \label{fig:scale}
 \end{center}
\end{figure}

\begin{figure}[h!]
    \centering
        \subfloat[]{\label{fig:scale_r:a}
        \includegraphics[width=3.5in]{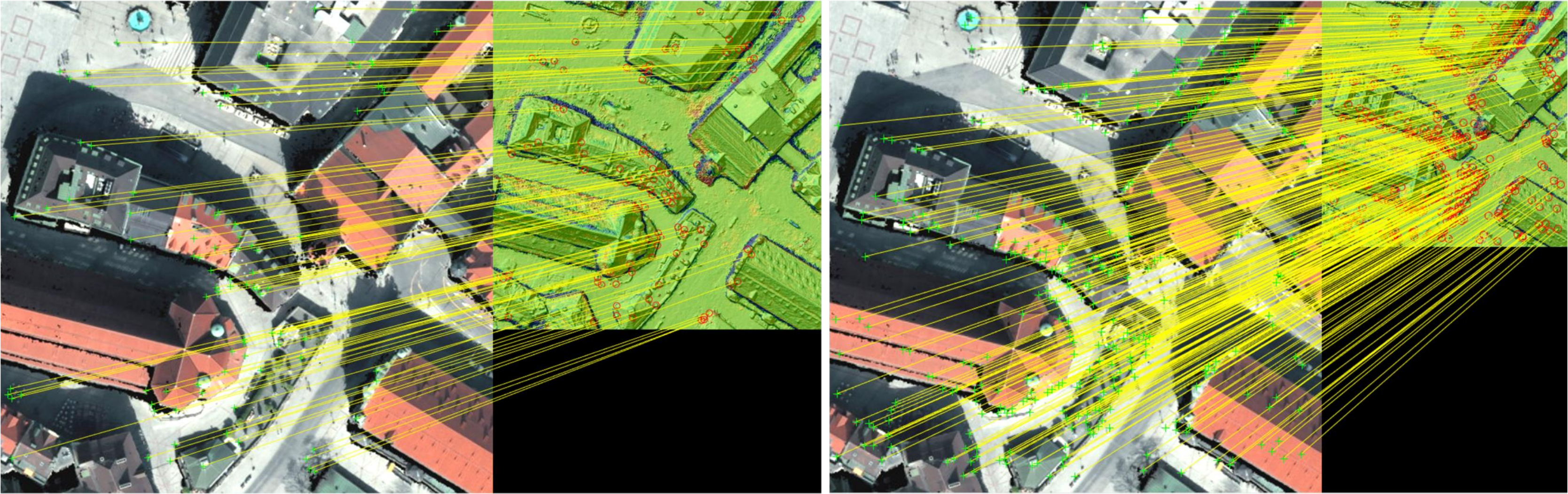}}
    \hfil
        \subfloat[]{\label{fig:scale_r:b}
        \includegraphics[width=3.5in]{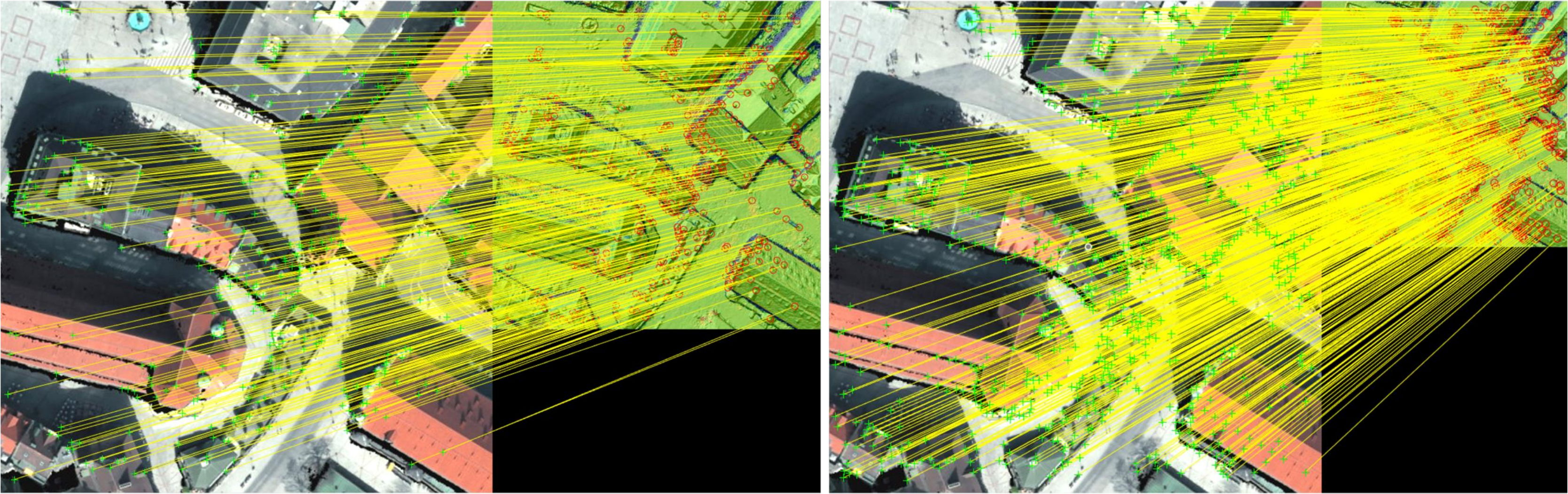}}
    \caption{Examples of feature matching results of scene (m) with scale differences. (a) Matching results by MS-HLMO with scale ratio of 1.5 and 2. (b) Matching results by MS-HLMO$^+$ with scale ratio of 1.5 and 2.}
    \label{fig:scale_r}
\end{figure}

To evaluate the scale invariance of MS-HLMO, the same strategy is used by fixing one image and scaling the other. The scaling ratios are from 1 to 2 with an interval of 0.1, providing a total of 11 image pairs in each scene. An example of the feature points matching results of MS-HLMO and MS-HLMO$^+$ are shown in Fig.\ref{fig:scale}-\ref{fig:scale_r}, respectively.

Even if there are scale differences between images with integer or non-integer proportions, the algorithm remains effective and obtains accurate matching. By observing the NCM curve, with the gradual change of the scale ratio from 1 to 2, the number of matching points gradually decreases, and reaches the lowest when the scale ratio was 1.5, and then gradually increases. This is because, in the experiment, the down-sampling ratio in building gaussian scale-space is set as 2. The interference caused by scale difference is well resisted when the sampling scale is integer multiple of 2. In other scale proportions, a significant difference of the receptive field in ground covers of the feature descriptor is caused between the image pair. So the local windows used for feature extraction between the two images are different in the receptive field. In practice, the sampling step in Gaussian pyramid is further refined to achieve better scale invariance.

This experiment demonstrates that MS-HLMO features are also robust to a certain extent when images are at scale proportion between two octaves in scale space and The proposed MS-HLMO has good robustness to scale difference of multi-source images, which plays an extremely positive role in the actual multi-source remote sensing image registration.

\begin{table*}[h!]
 \centering
 \setlength{\tabcolsep}{1.2mm}
 \caption{NCMs of the 17 remote sensing scenes comparison by eight registration methods.}
 \label{tbl:result}
 \begin{tabular}{lcccccccccccccccccccc}
  \toprule
  \multirow{2}{*}{Method} & \multicolumn{20}{c}{Scene}\\
    \cline{2-21} &   a1 &  a2 &  a3 &  a4 &   b &   c &    d &   e &   f &   g &   h &   i &   j &   k &   l &   m &   n &   o &   p &   q \\
  \midrule
   SIFT          &  794 &  30 &  25 &  —— &  —— & 145 &   86 &  —— &  —— &  —— &  —— &  —— &  —— &  —— &  —— &  —— &  —— &  15 &  14 & 256 \\
   SAR-SIFT      &   16 &   6 &  —— &  —— &  —— &  66 &    4 &  —— &  —— &  —— &  —— &  —— &  —— &  —— &  —— &  —— &  —— &   9 &  36 & 262 \\
   PSO-SIFT      &    3 &  —— &  —— &  —— &  —— &  —— &   —— & 171 &  13 &  —— &  —— &  —— &   9 &   6 &  10 &  —— &  —— &  19 &  24 & 259 \\
   PIIFD         &   —— &  —— &  —— &  —— &  —— &  —— &   —— &  53 &  —— &  —— &  —— &  —— &  —— &  —— &  —— &  —— &  —— &  —— &  —— & 493 \\
   MS-PIIFD      & 1105 &  —— & 481 &  —— & 444 & 145 &  999 & 460 &  78 &  —— &  —— &  25 &   5 & 339 &  30 &  74 & 165 &  —— & 322 & 860 \\
   RIFT$^+$    &   —— &  21 &  —— &  34 &  —— &  —— &   —— &  —— &  —— &  —— &  —— &  39 & 176 & 109 & 188 & 189 &  96 & 179 &  —— &  —— \\
   MS-HLMO       & 1126 & 112 & 865 & 119 & 603 & 394 & 1030 & 689 & 161 & 130 & 231 &  79 & 149 & 511 & 420 & 374 & 161 &  97 & 432 & 883 \\
   MS-HLMO$^+$ & 1194 & 321 & 996 & 149 & 984 & 564 & 1163 &  —— & 693 & 162 & 337 & 145 & 761 & 789 & 861 & 846 & 273 & 451 &  —— &  —— \\
  \bottomrule
 \end{tabular}
\end{table*}

\begin{figure*}[h!]
    \begin{center}
        \subfloat[]{\label{fig:matching:a}
        \includegraphics[height=1.0in]{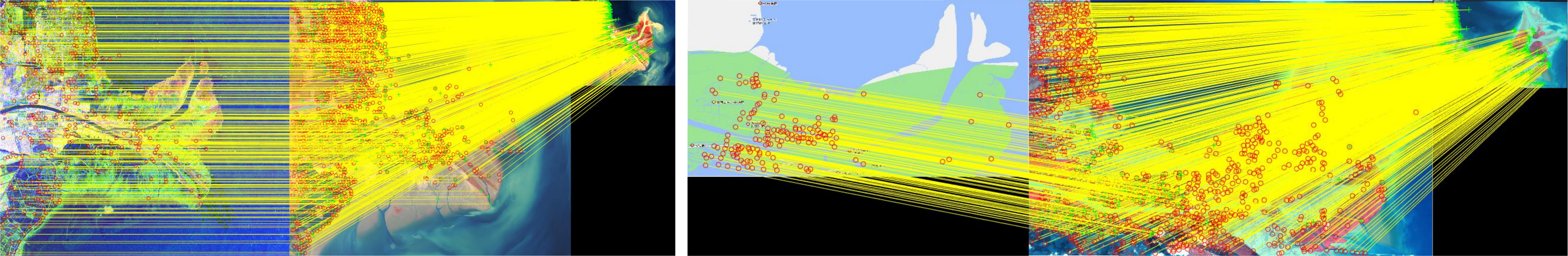}}
        \subfloat[]{\label{fig:matching:b}
        \includegraphics[height=1.0in]{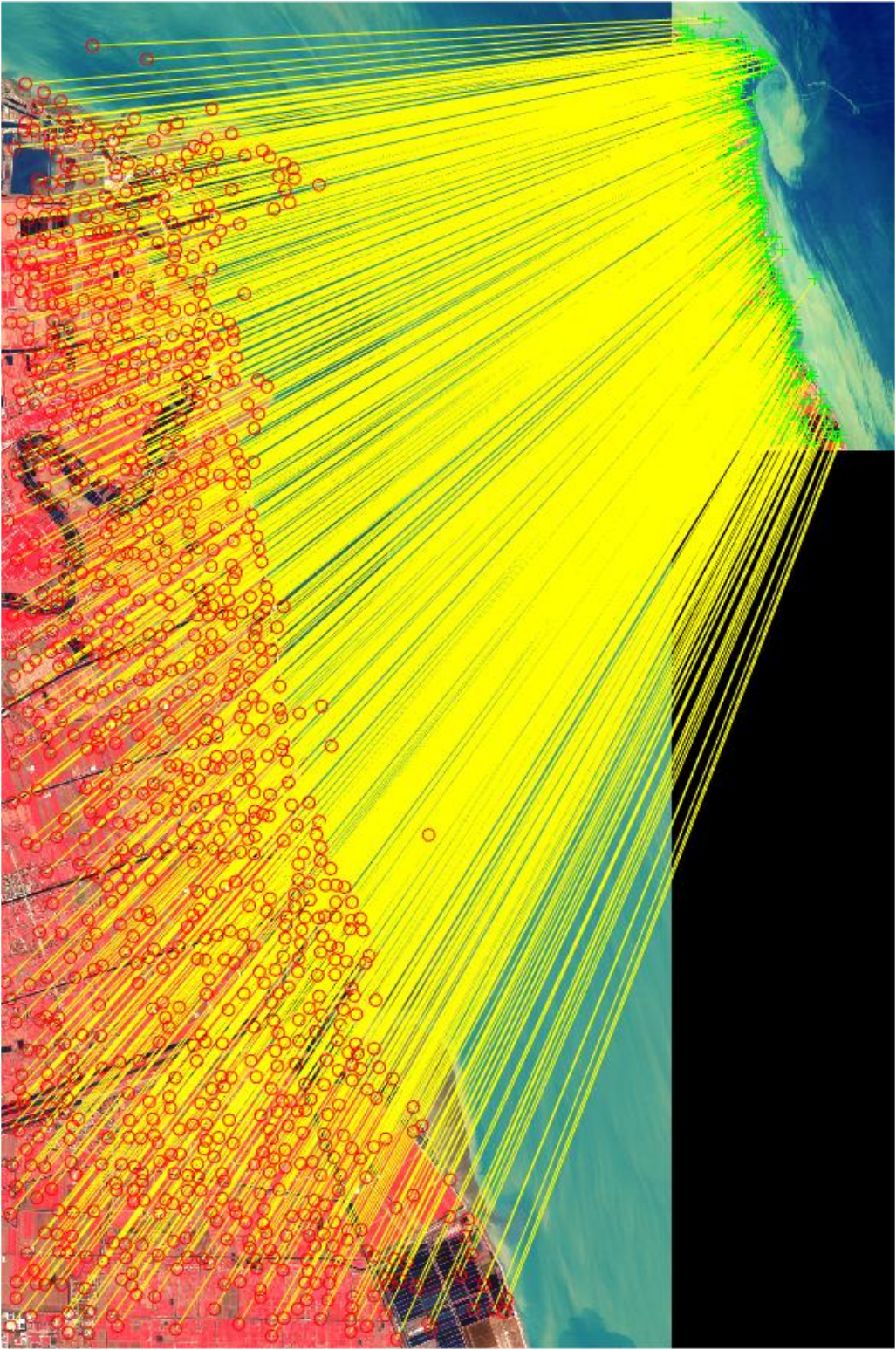}}

        \subfloat[]{\label{fig:matching:c}
        \includegraphics[height=0.8in]{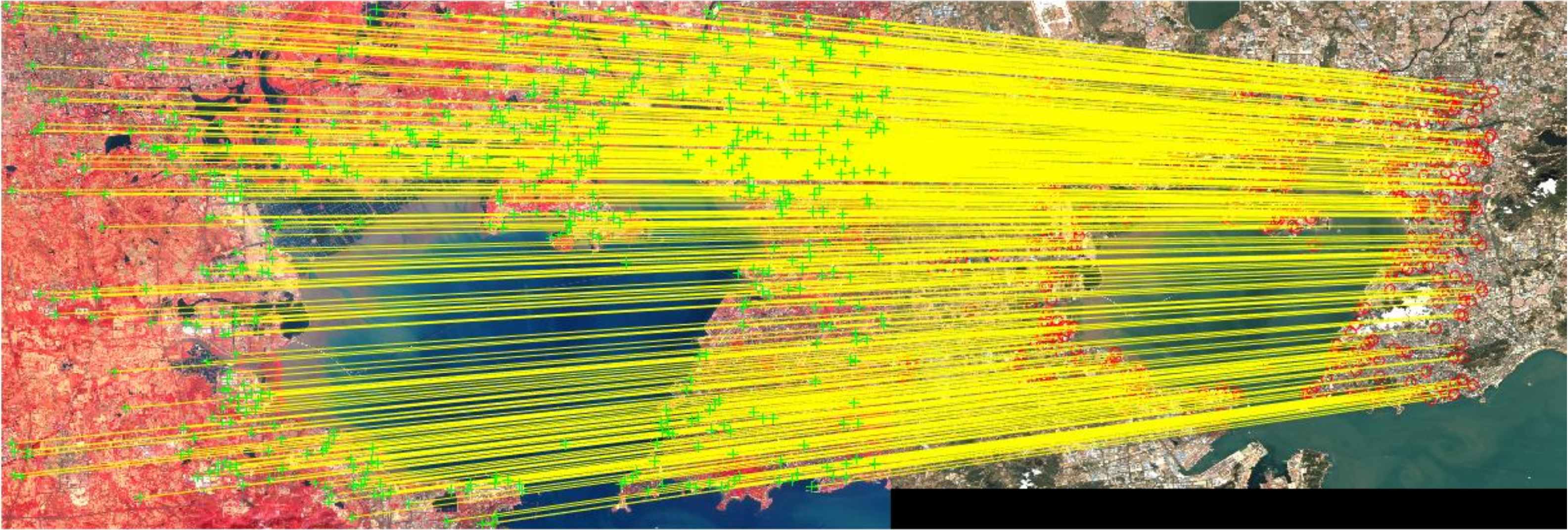}}
        \subfloat[]{\label{fig:matching:d}
        \includegraphics[height=0.8in]{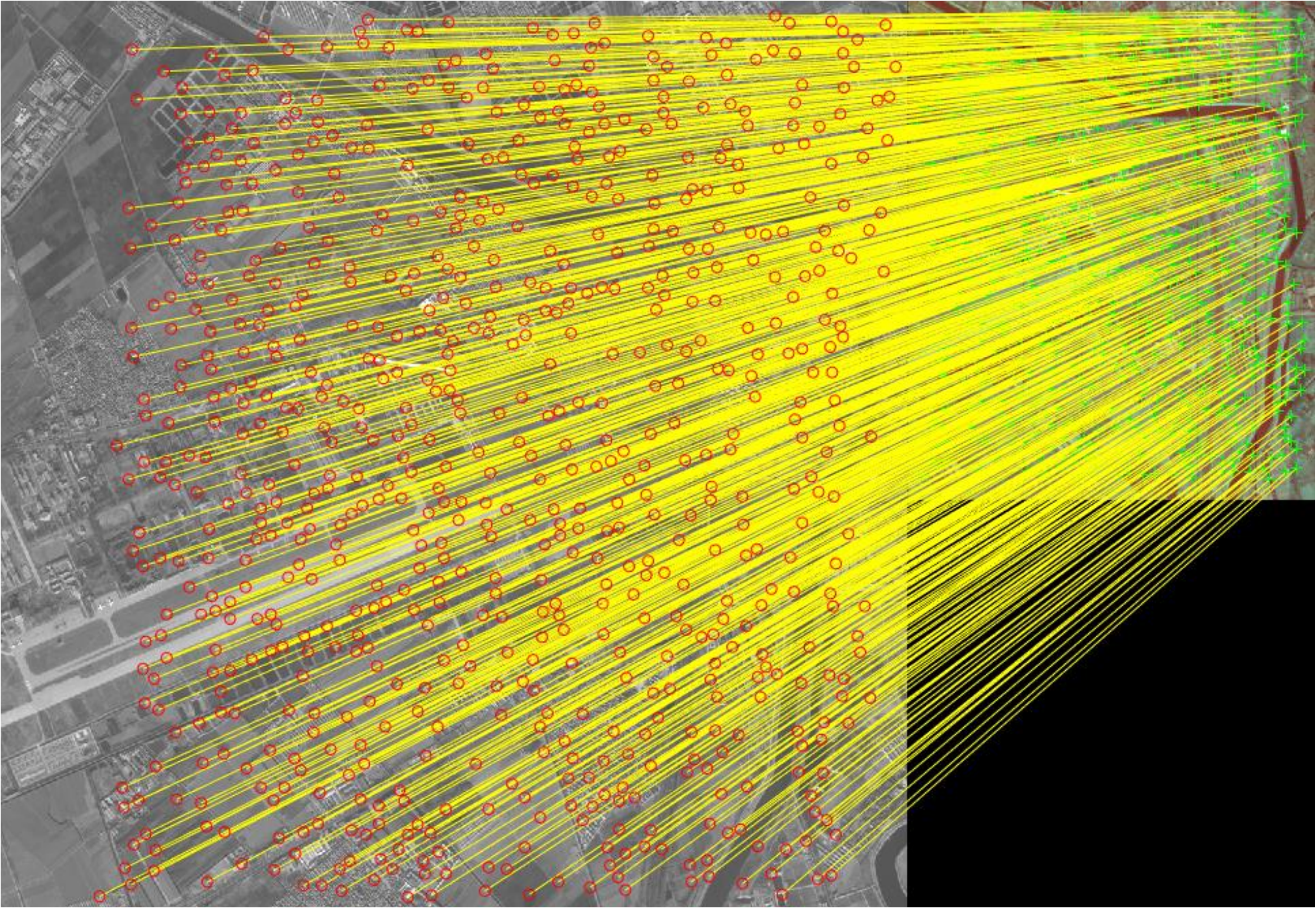}}
        \subfloat[]{\label{fig:matching:e}
        \includegraphics[height=0.8in]{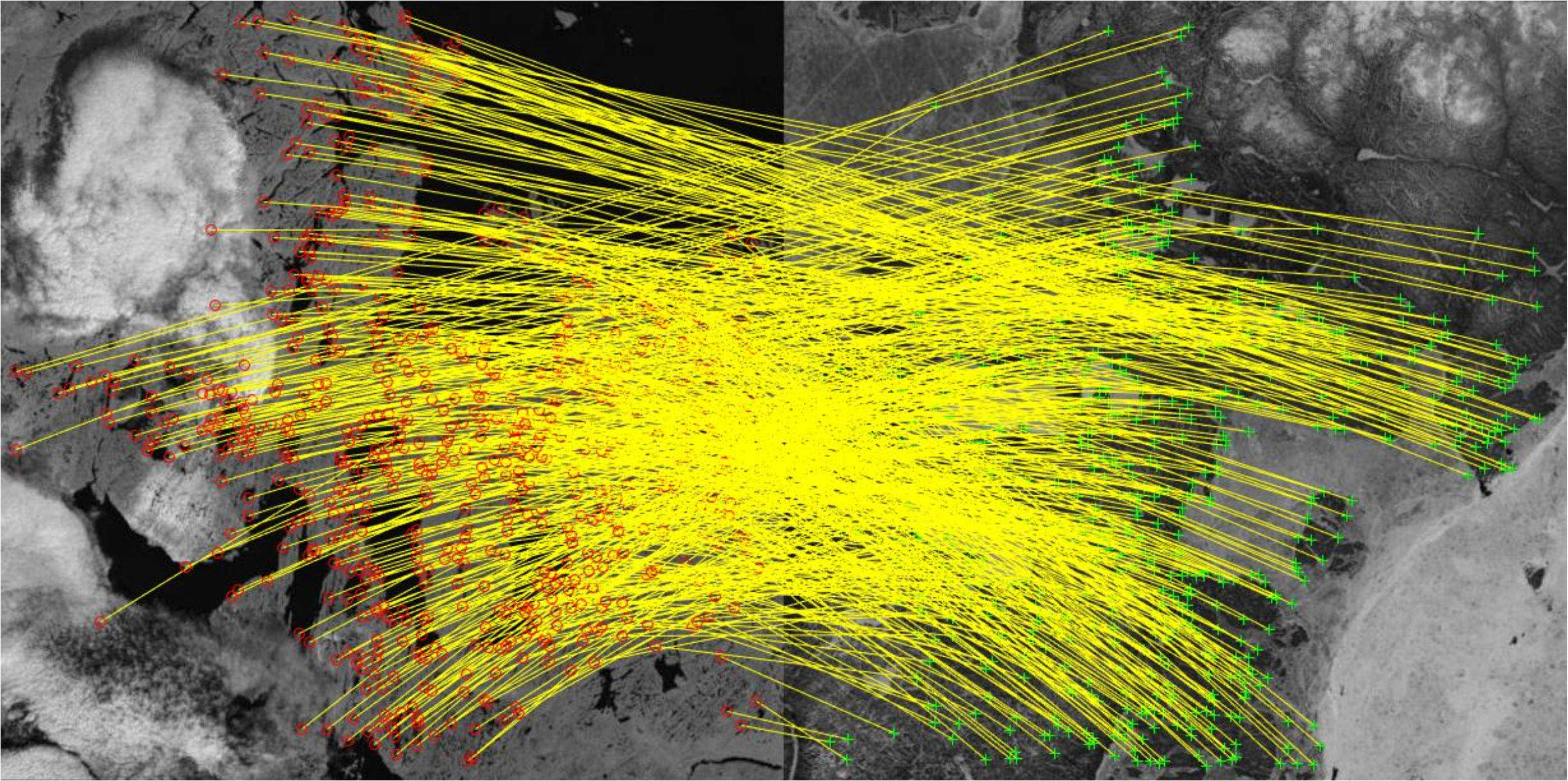}}
        \subfloat[]{\label{fig:matching:f}
        \includegraphics[height=0.8in]{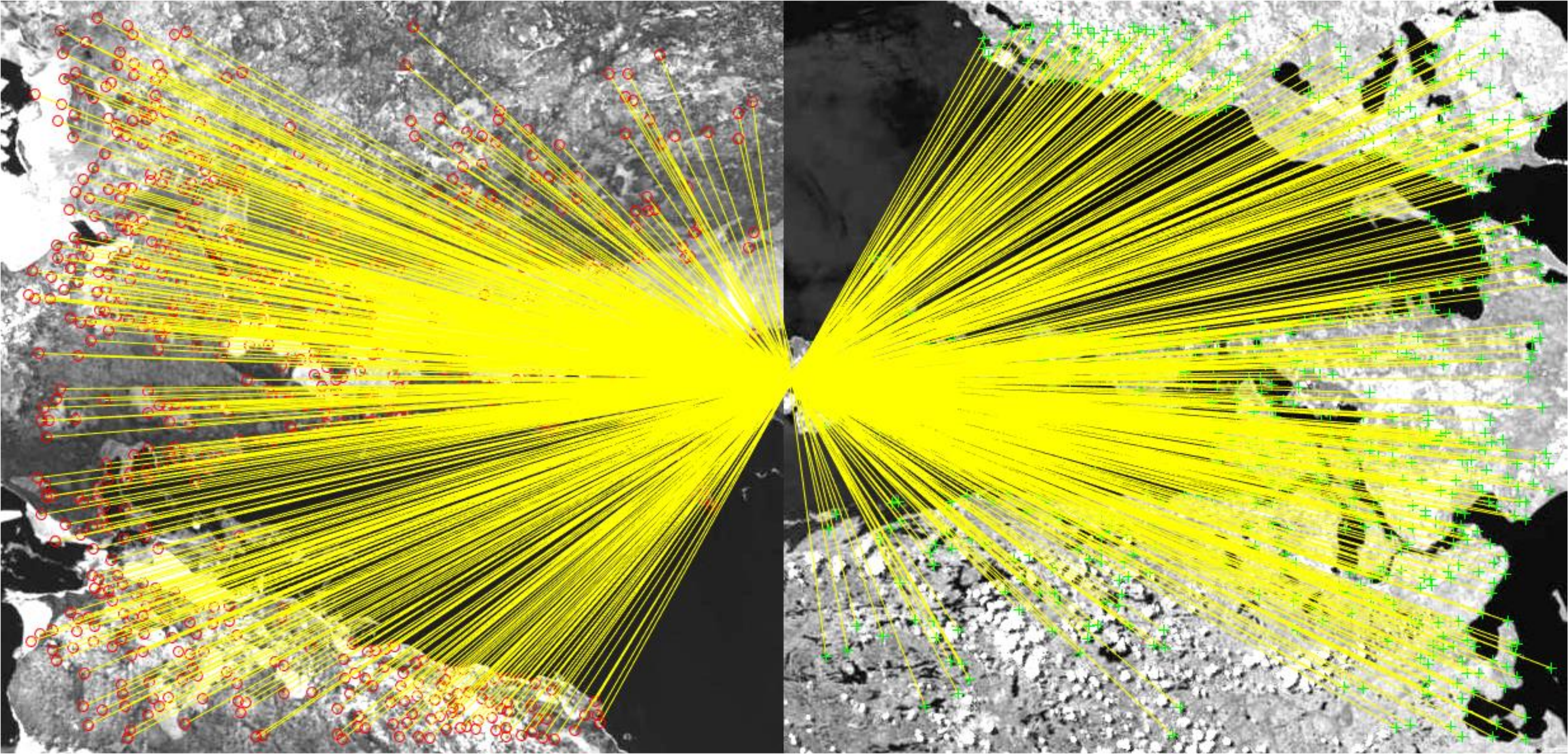}}

        \subfloat[]{\label{fig:matching:g}
        \includegraphics[height=1.0in]{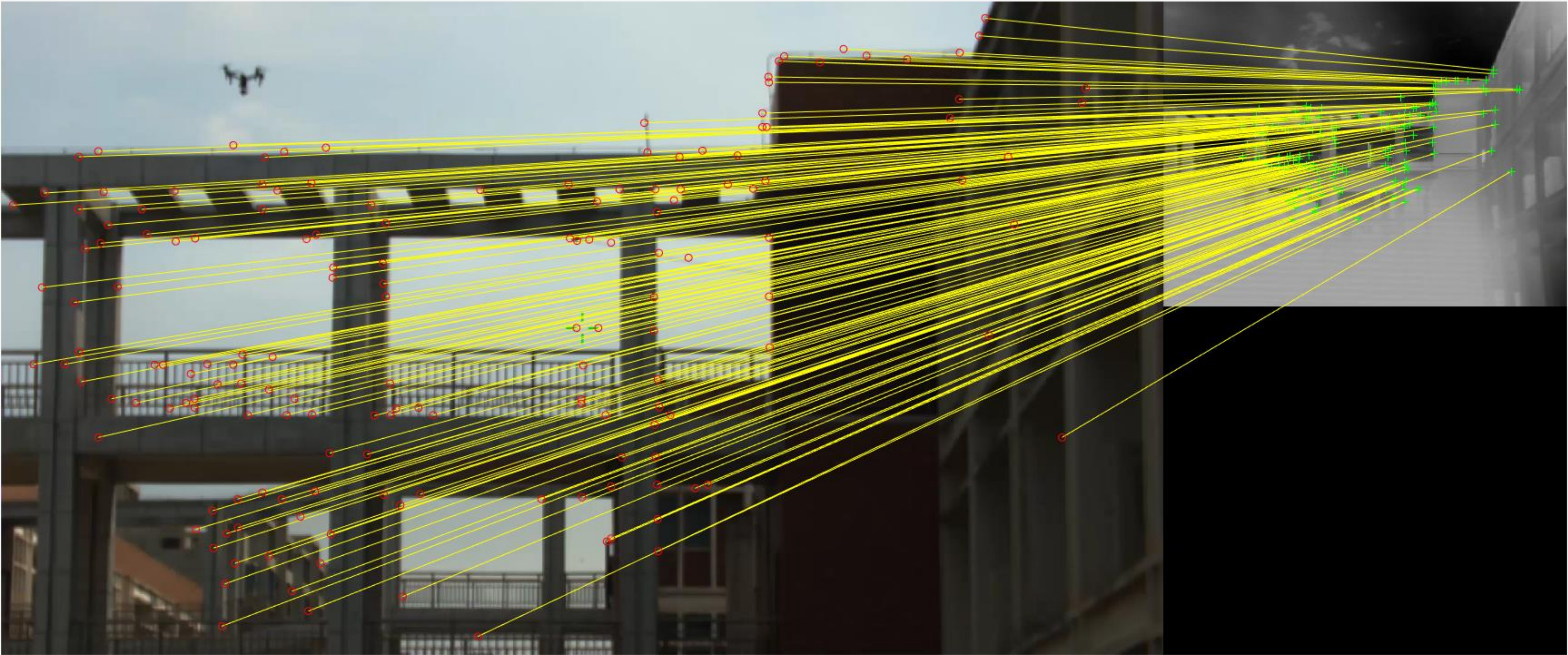}}
        \subfloat[]{\label{fig:matching:h}
        \includegraphics[height=1.0in]{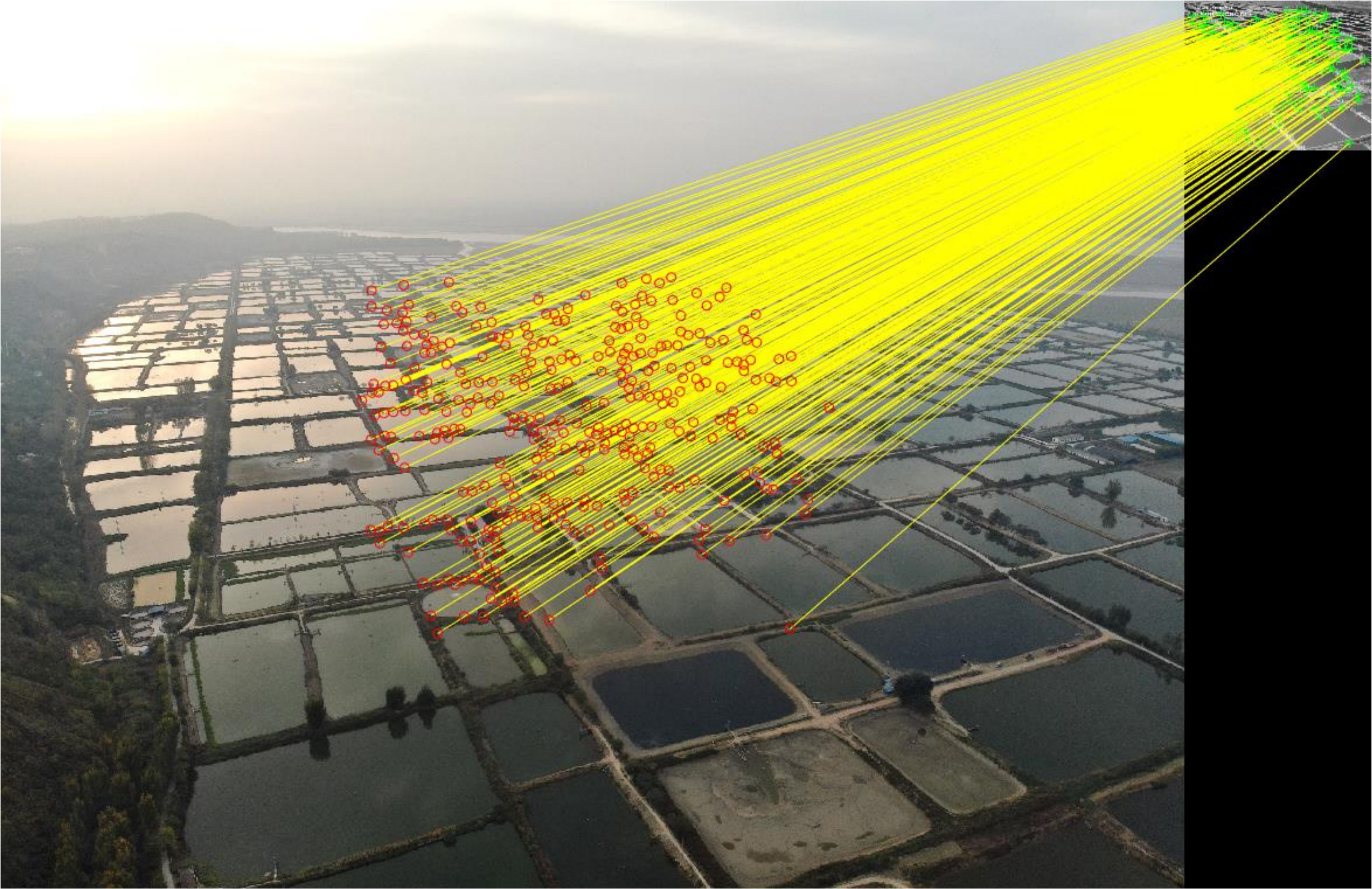}}
        \subfloat[]{\label{fig:matching:i}
        \includegraphics[height=1.0in]{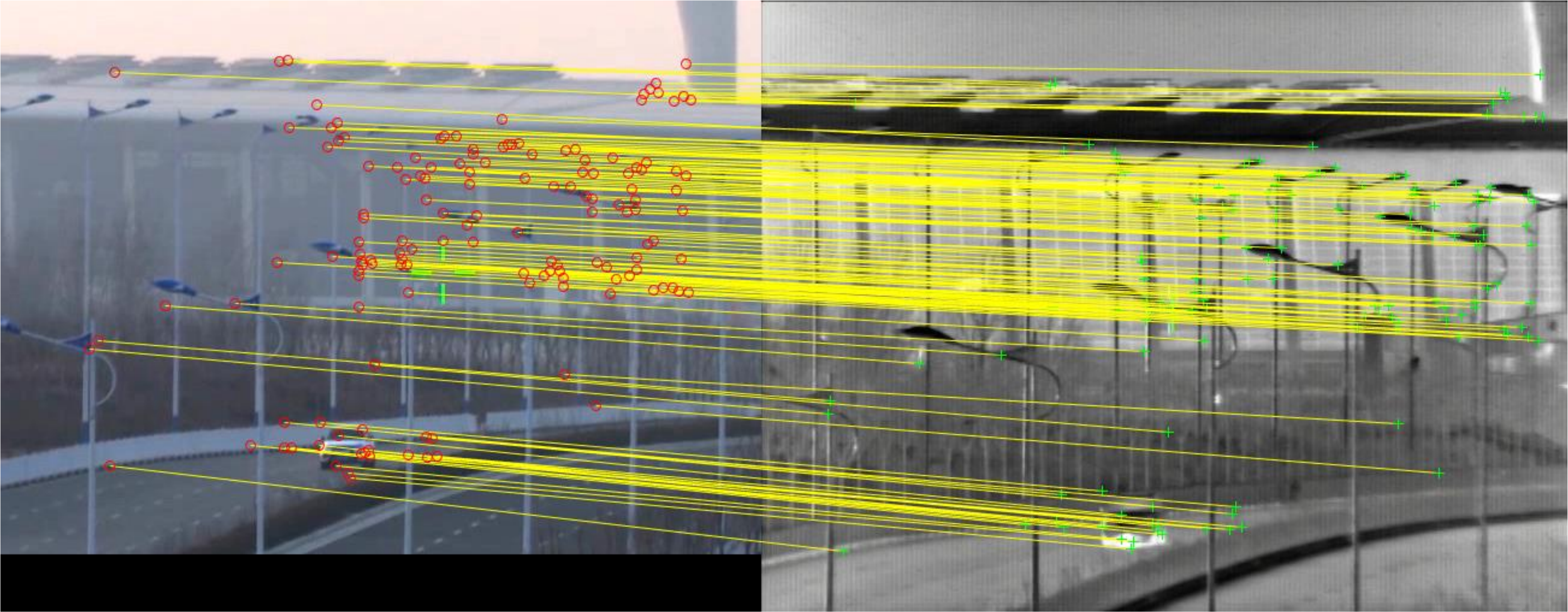}}

        \subfloat[]{\label{fig:matching:j}
        \includegraphics[height=0.8in]{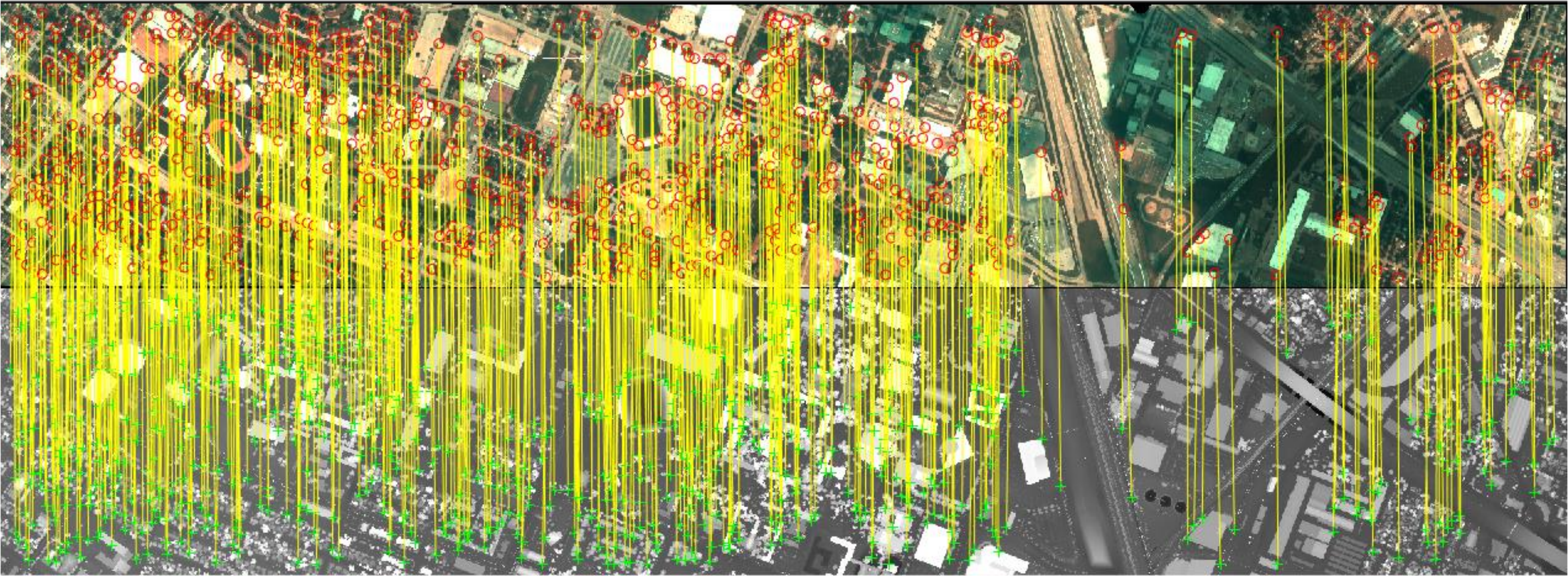}}
        \subfloat[]{\label{fig:matching:k}
        \includegraphics[height=0.8in]{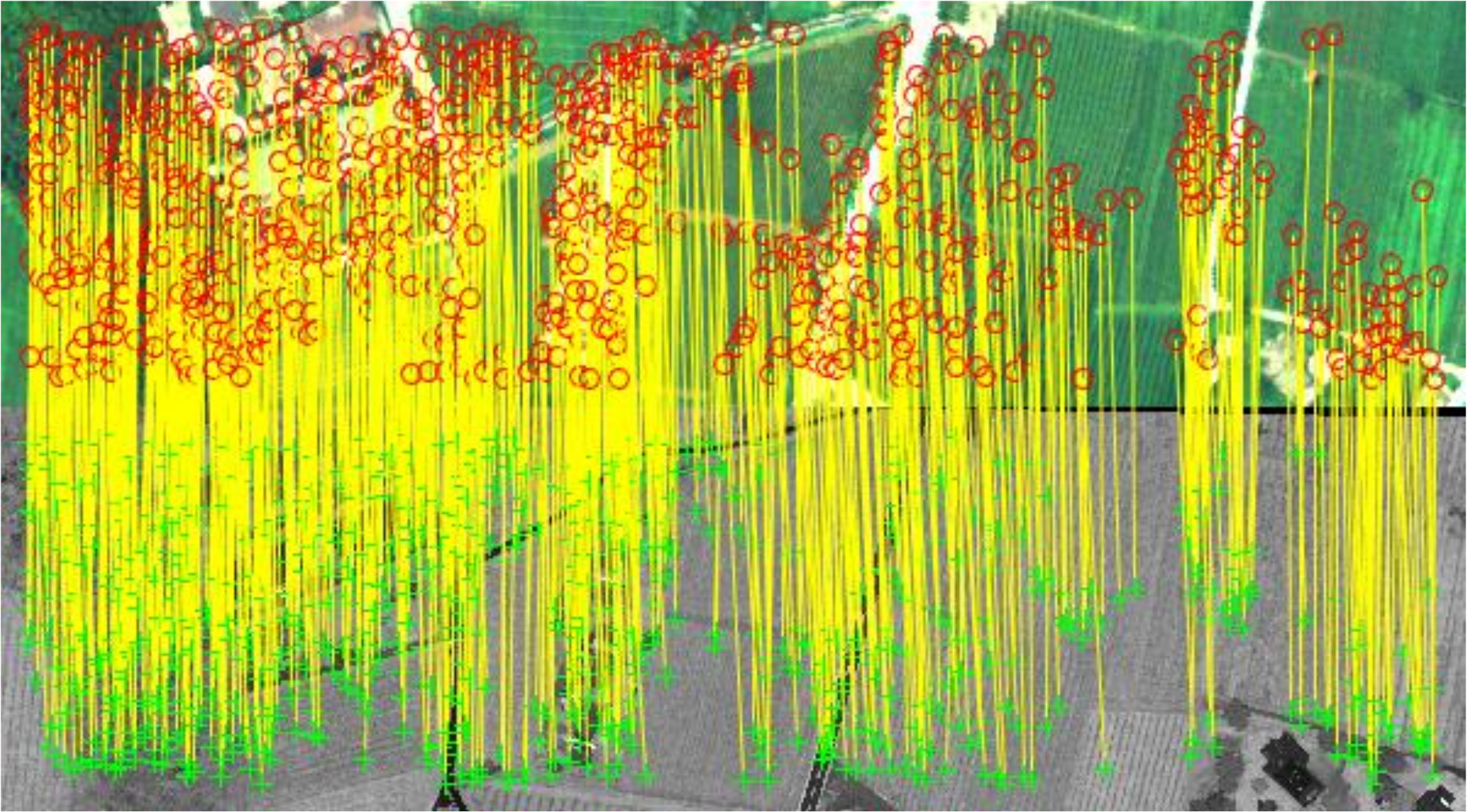}}
        \subfloat[]{\label{fig:matching:l}
        \includegraphics[height=0.8in]{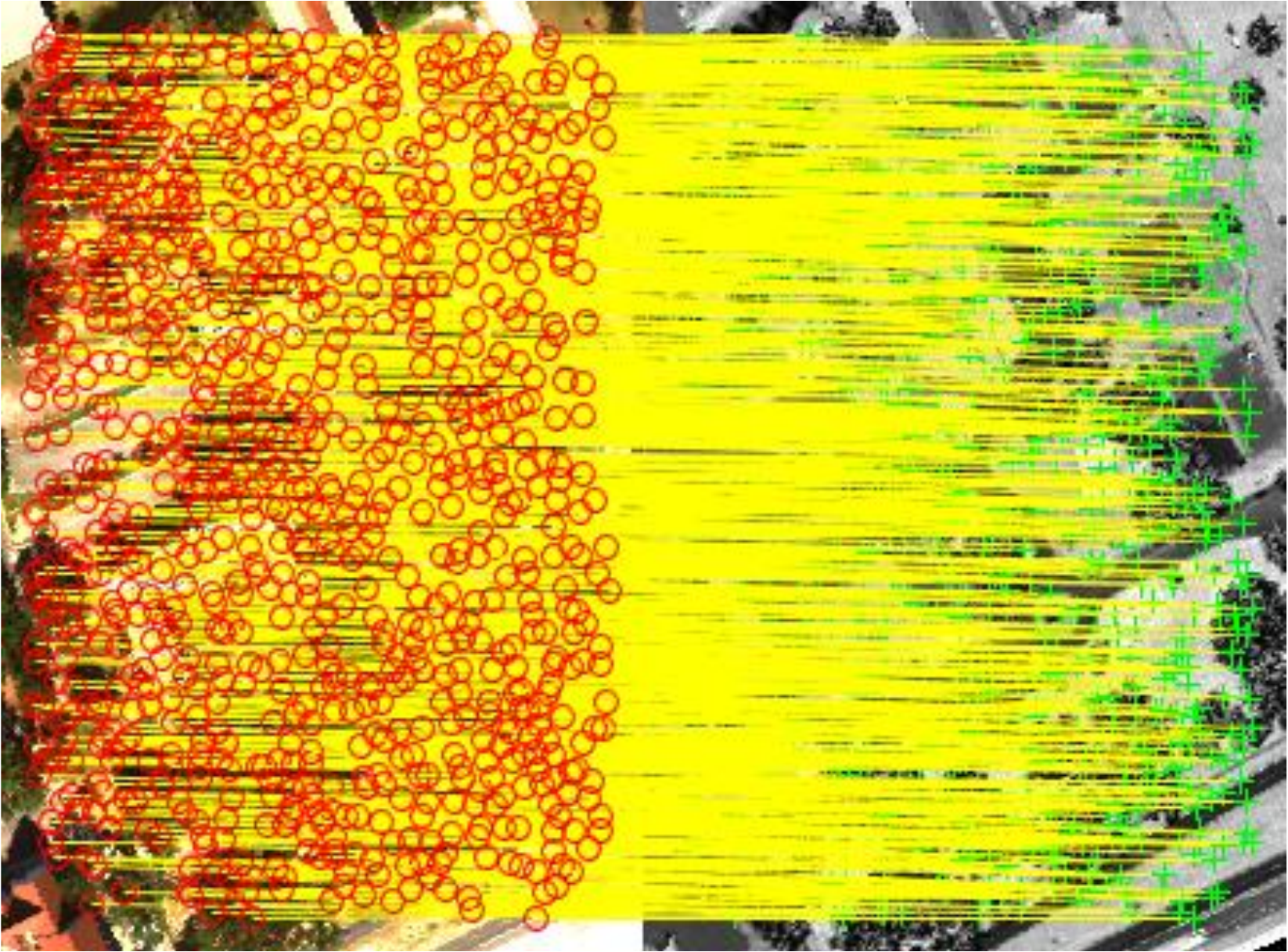}}
        \subfloat[]{\label{fig:matching:m}
        \includegraphics[height=0.8in]{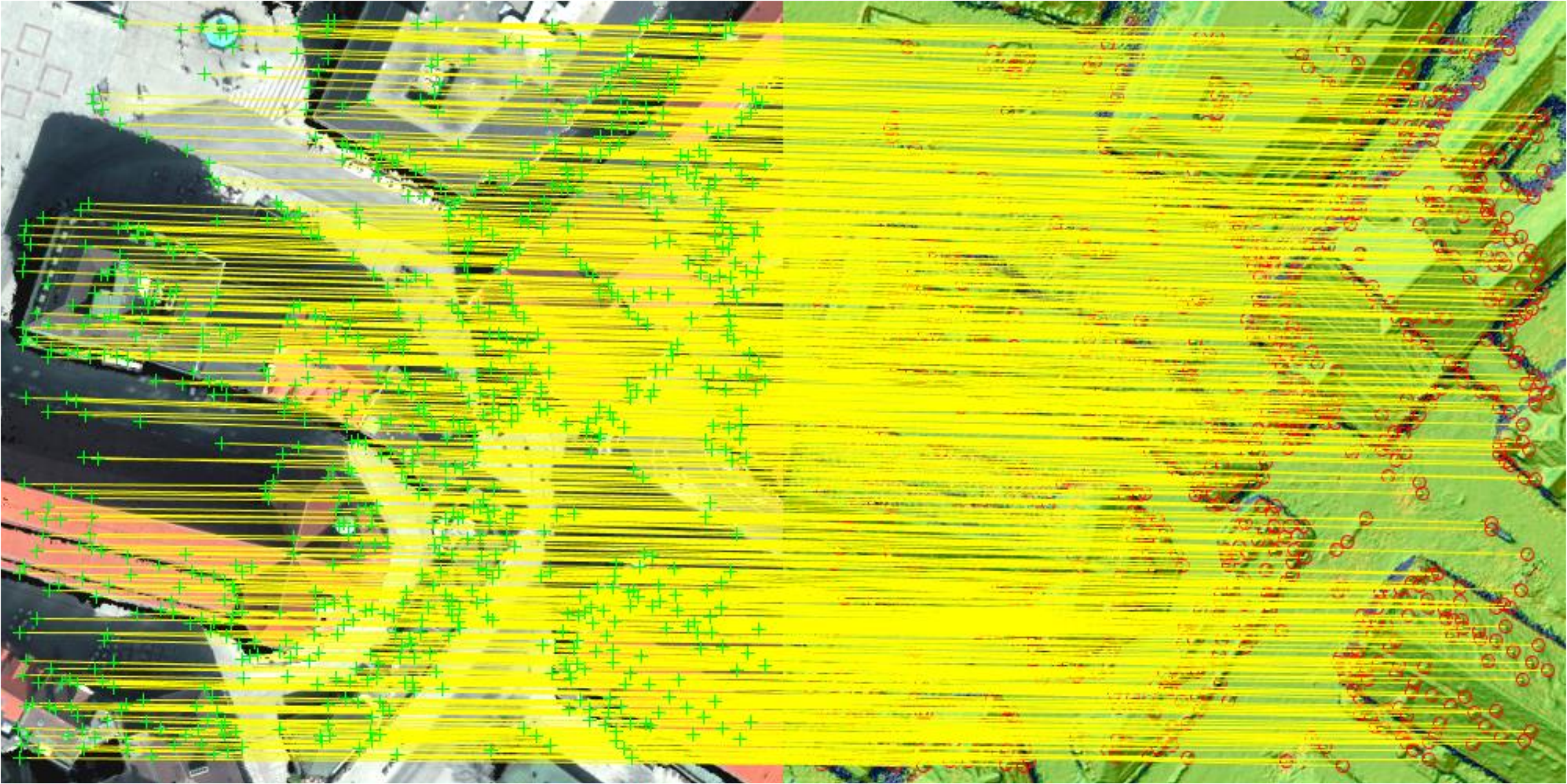}}

        \subfloat[]{\label{fig:matching:n}
        \includegraphics[height=0.8in]{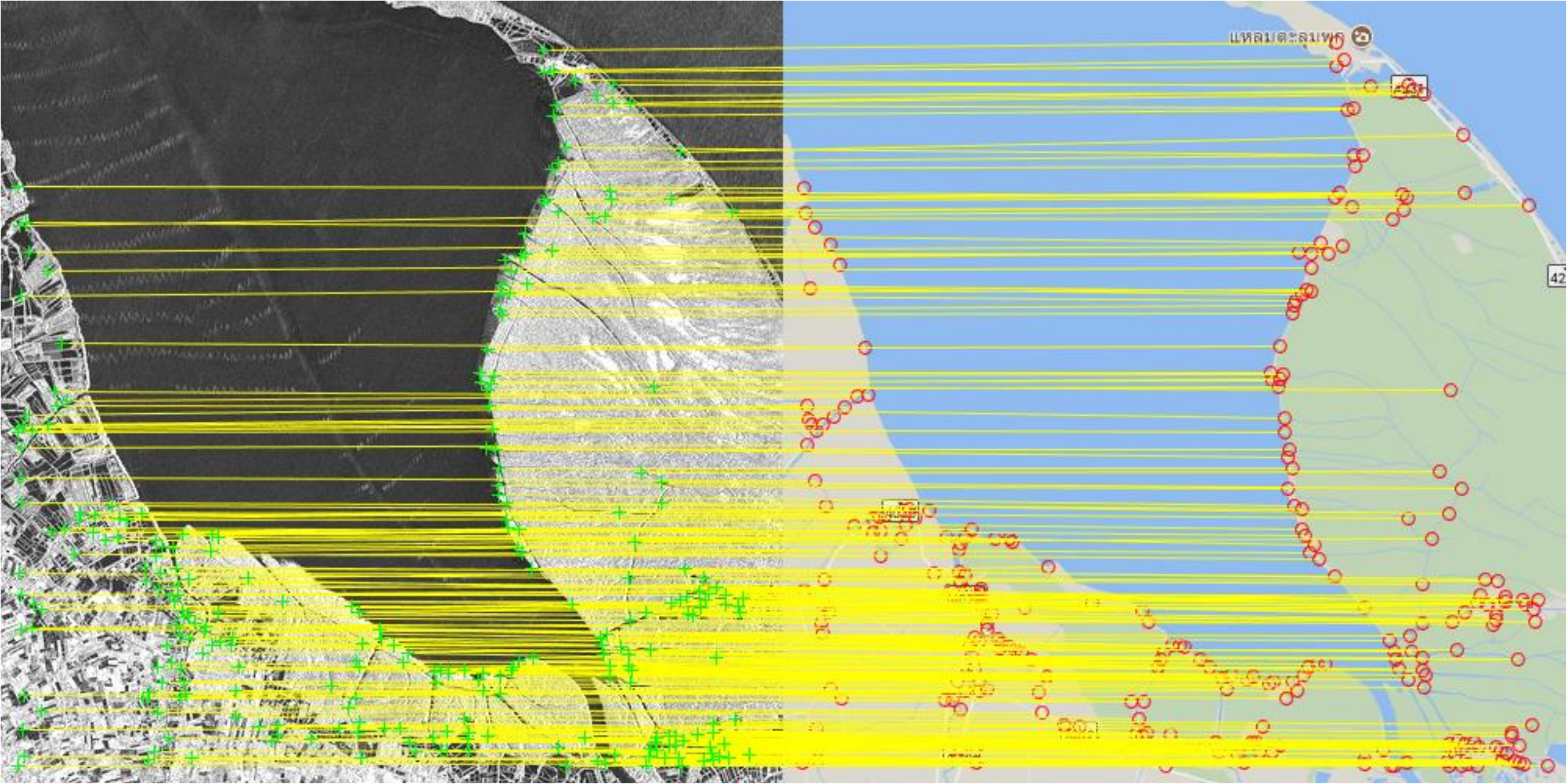}}
        \subfloat[]{\label{fig:matching:o}
        \includegraphics[height=0.8in]{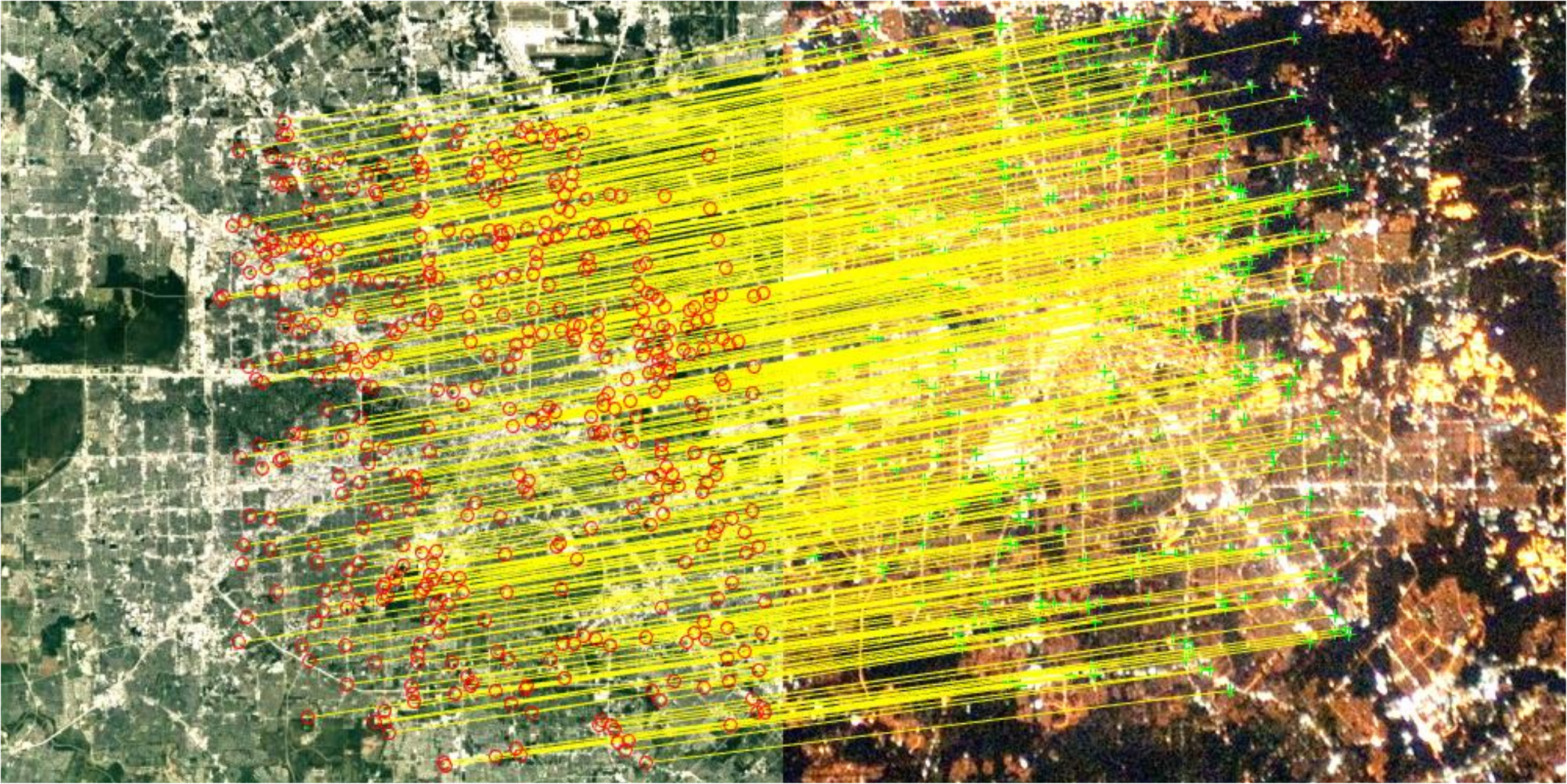}}
        \subfloat[]{\label{fig:matching:p}
        \includegraphics[height=0.8in]{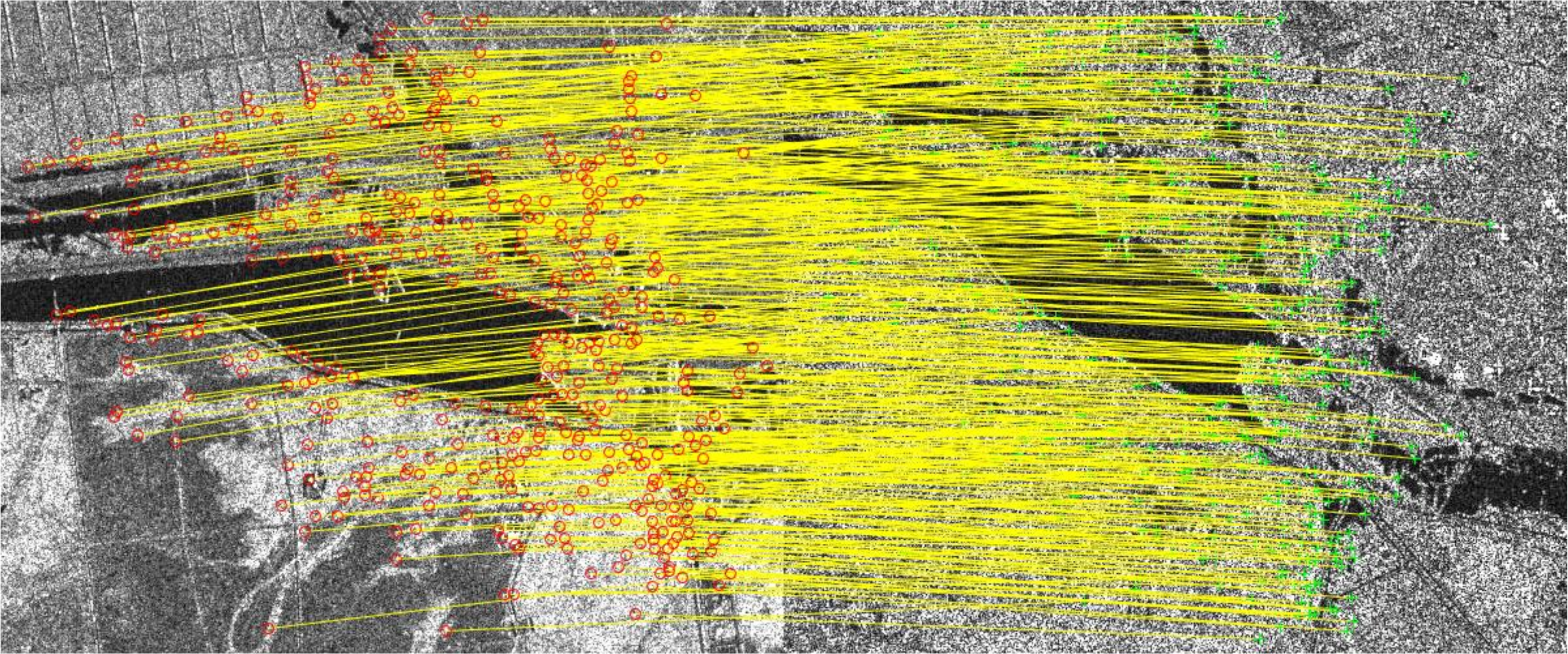}}
        \subfloat[]{\label{fig:matching:q}
        \includegraphics[height=0.8in]{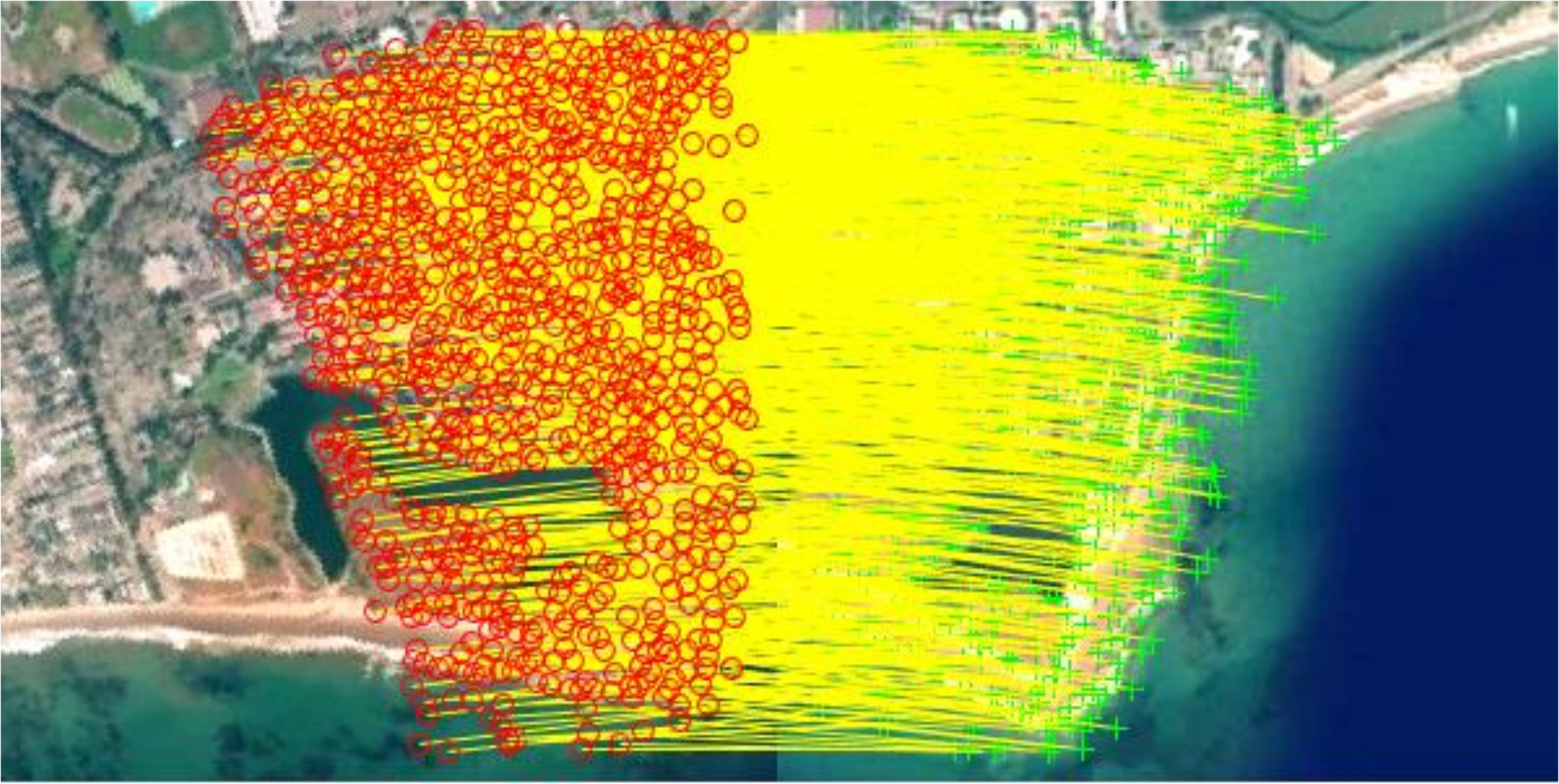}}
    \end{center}
    \caption{Feature points matching results of the 17 remote sensing scenes by the proposed MS-HLMO. (a) a2:SAR-MSI, a1:MSI-HSI, a4:map-MSI, and a3:MSI-HSI (multi-temporal). (b) MSI-HSI. (c) MSI-MSI. (d) PAN-HSI. (e) visible-infrared. (f) visible-infrared. (g) visible-infrared. (h) visible-infrared. (i) visible-infrared. (j) HSI-LiDAR. (k) HSI-LiDAR. (l) HSI-LiDAR. (m) visible-depth. (n) visible-map. (o) day-night (visible). (p) multi-source SAR. (q) single-source visible.}
    \label{fig:matching}
\end{figure*}

\subsection{Registration Performance}
\label{ssec:final}

\begin{figure*}[!h]
    \begin{center}
        \subfloat[]{\label{fig:checker:a}
        \includegraphics[height=0.85in]{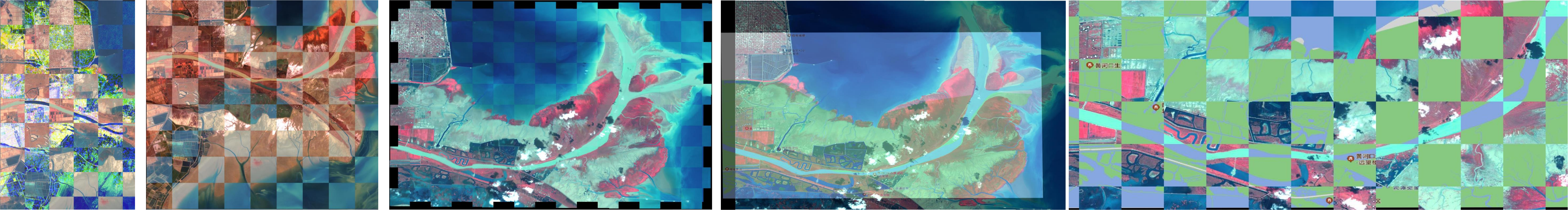}}
        \subfloat[]{\label{fig:checker:b}
        \includegraphics[height=1.1in]{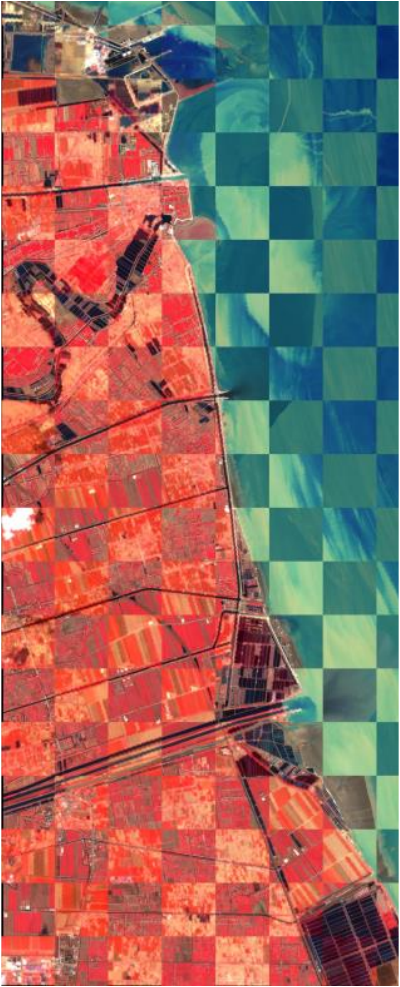}}

        \subfloat[]{\label{fig:checker:c}
        \includegraphics[height=0.8in]{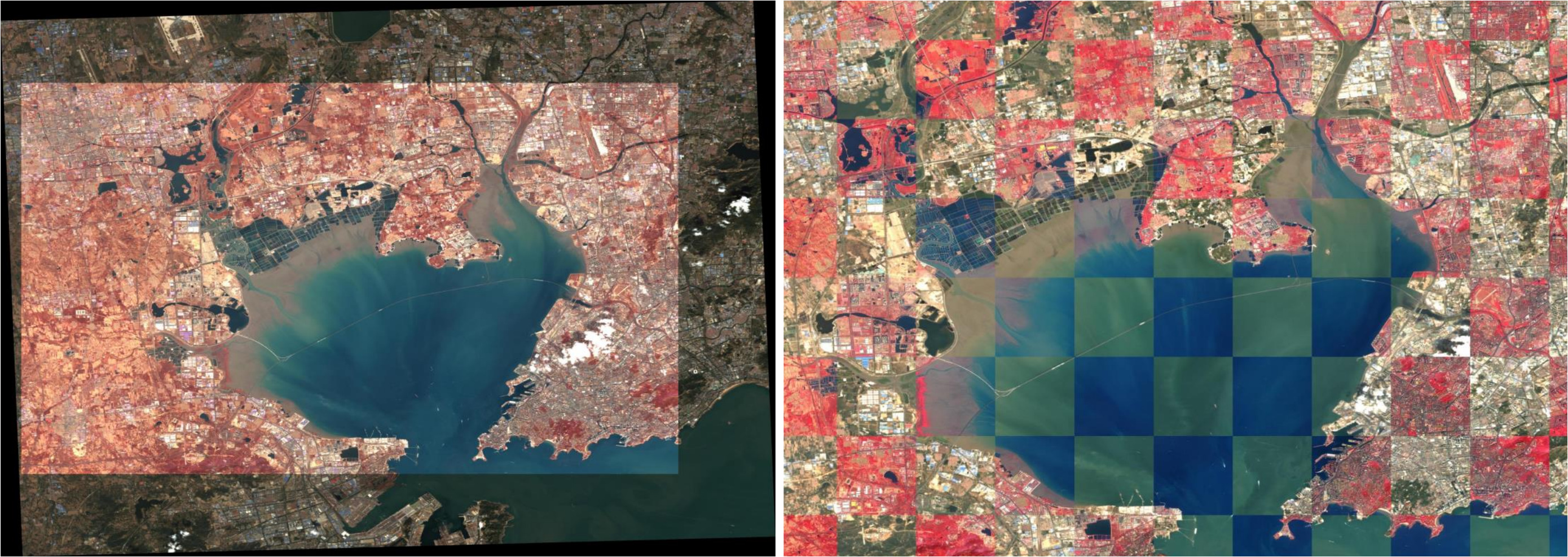}}
        \subfloat[]{\label{fig:checker:d}
        \includegraphics[height=0.8in]{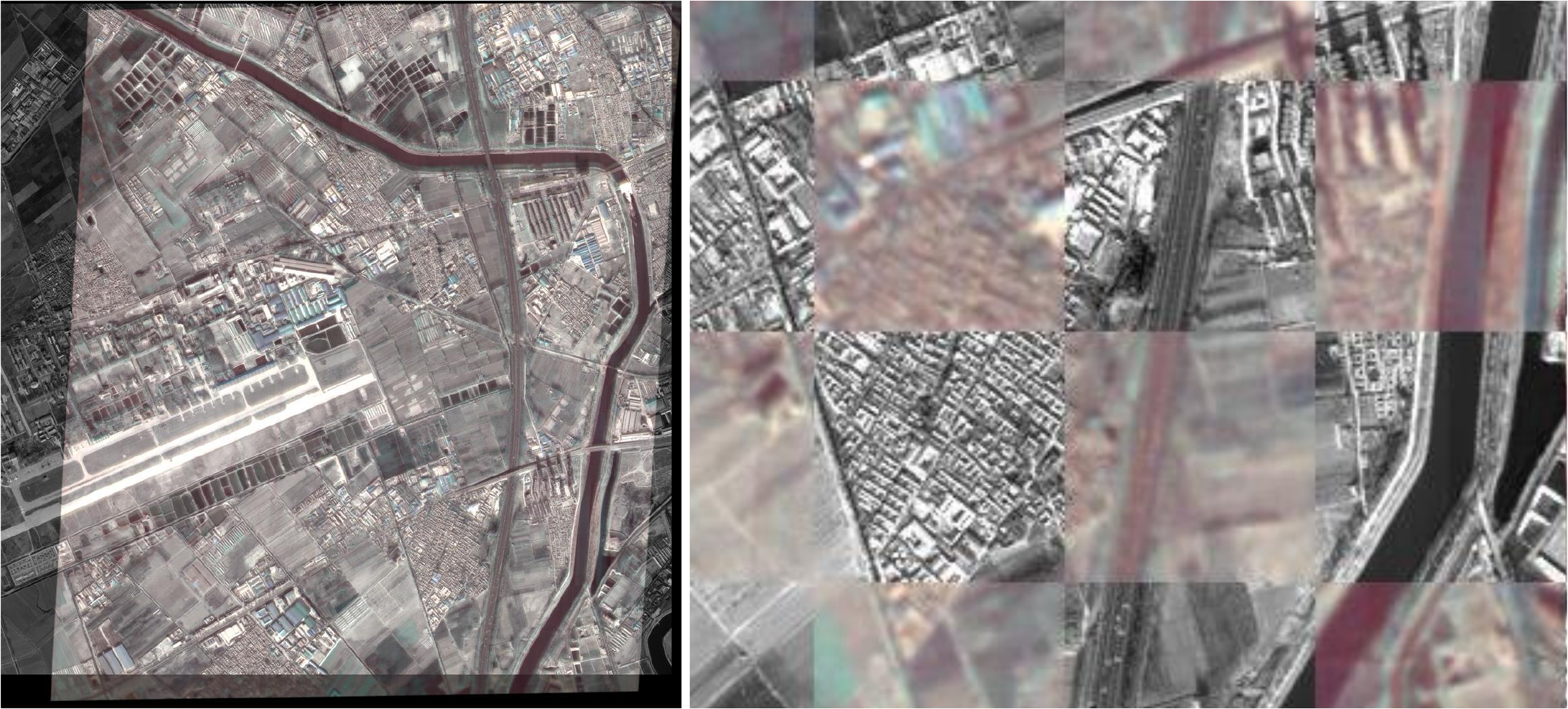}}
        \subfloat[]{\label{fig:checker:e}
        \includegraphics[height=0.8in]{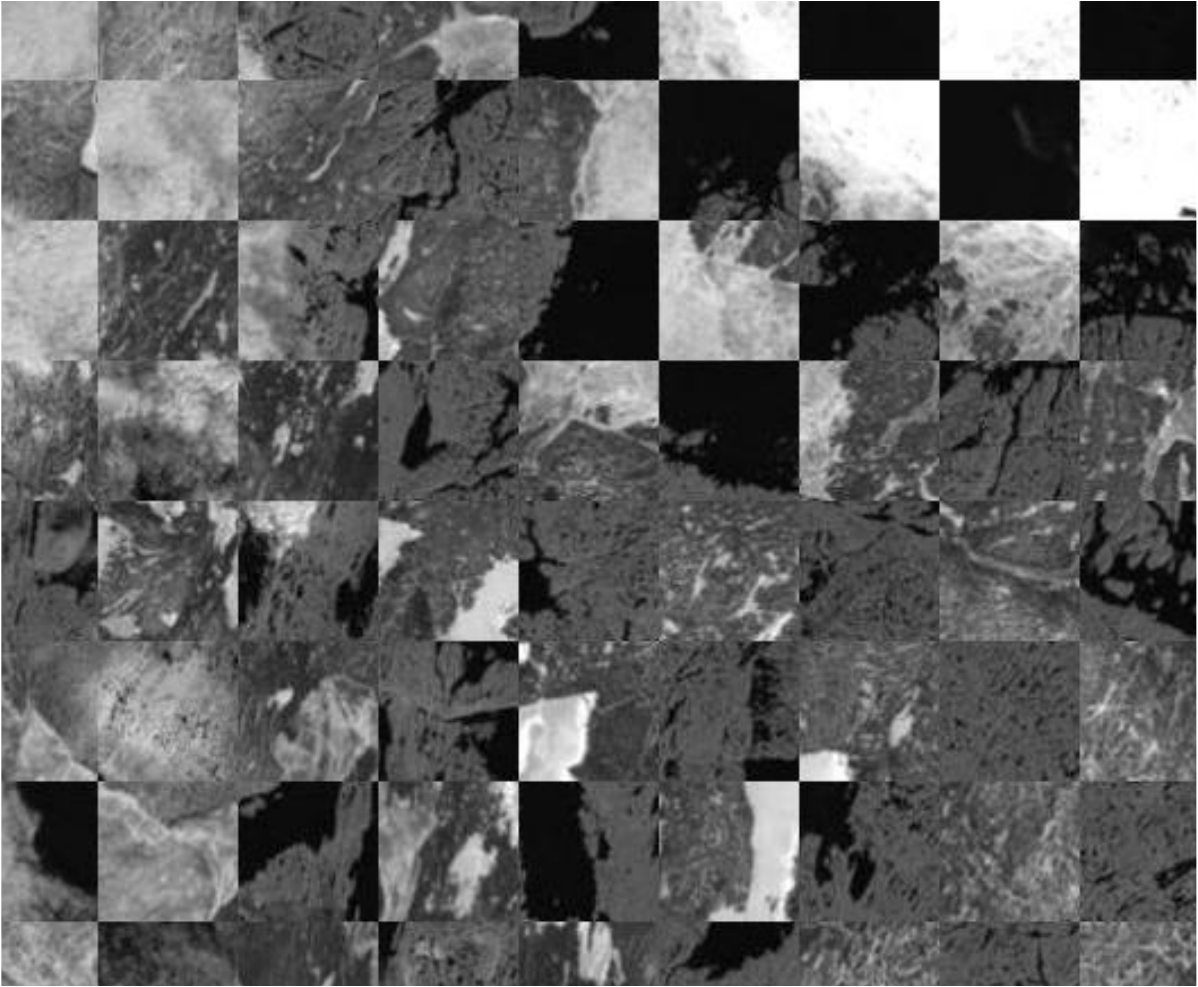}}
        \subfloat[]{\label{fig:checker:f}
        \includegraphics[height=0.8in]{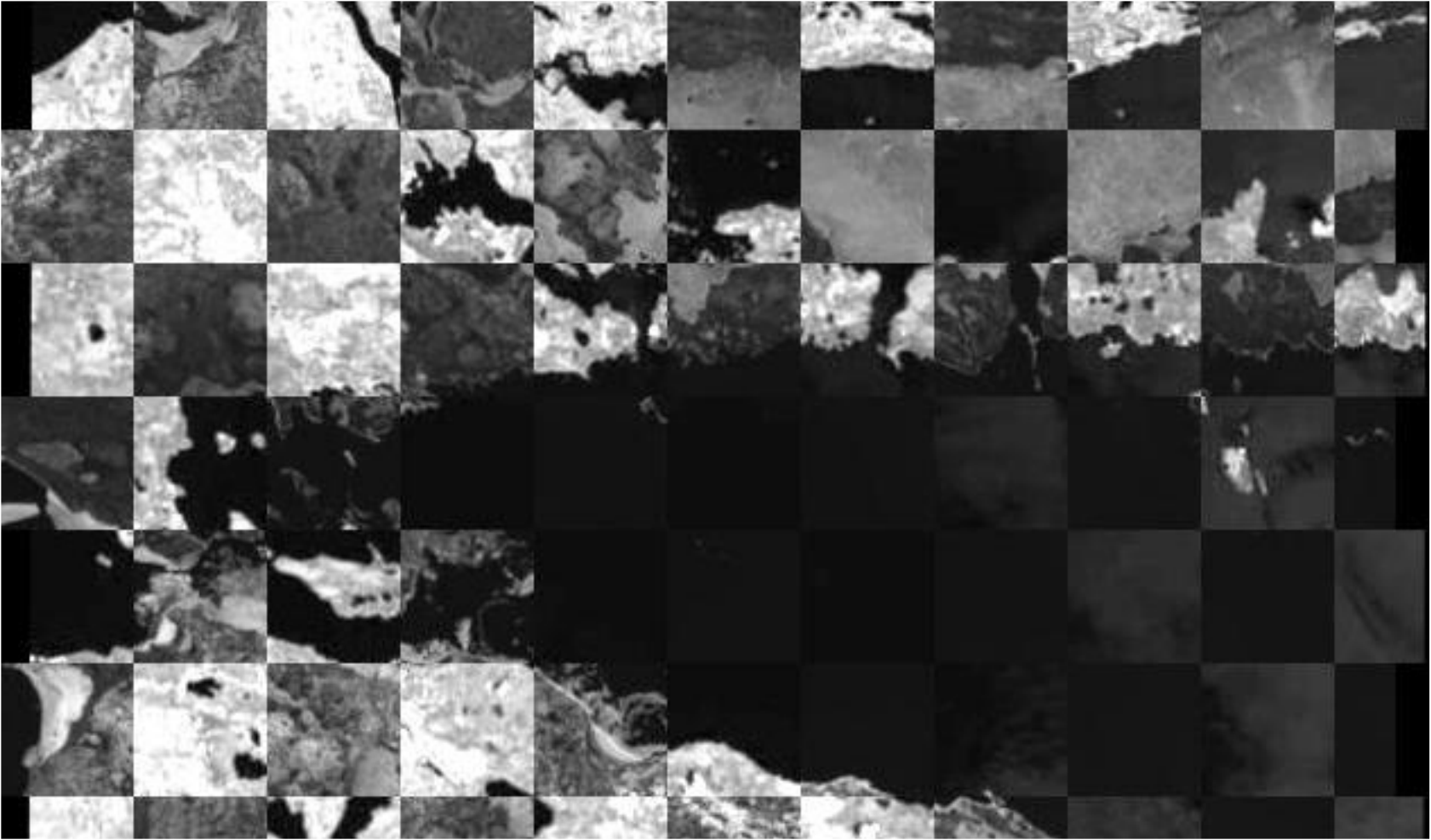}}

        \subfloat[]{\label{fig:checker:g}
        \includegraphics[height=0.8in]{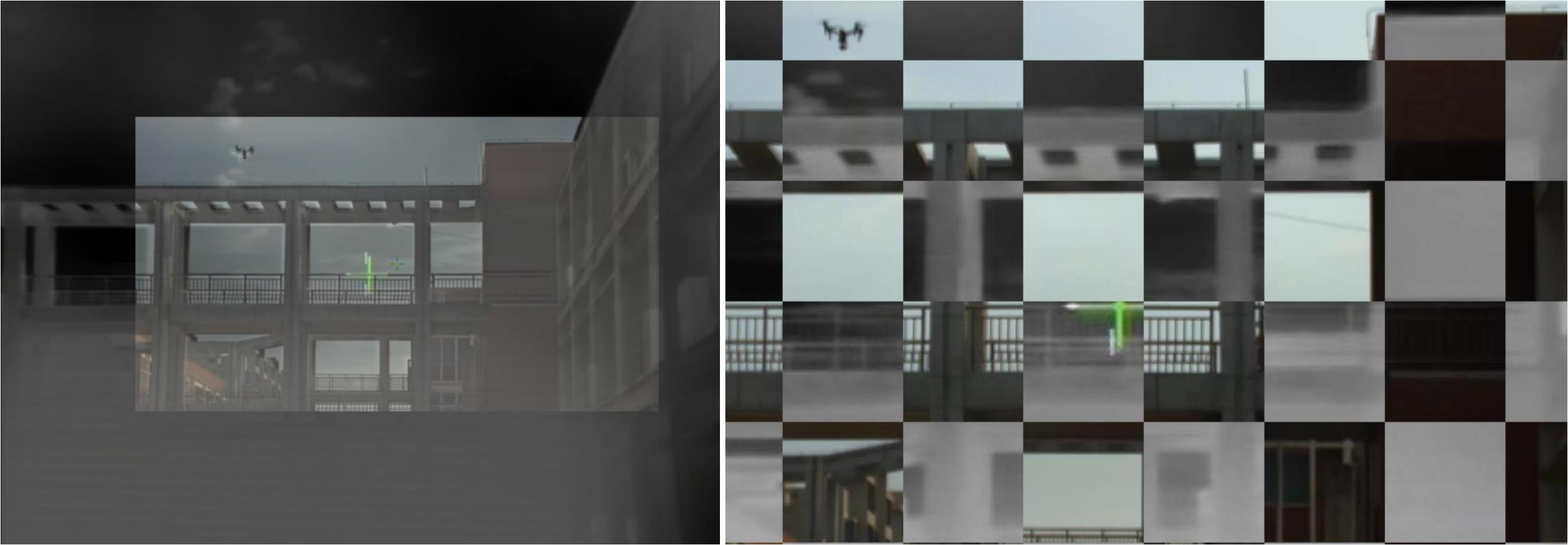}}
        \subfloat[]{\label{fig:checker:h}
        \includegraphics[height=0.8in]{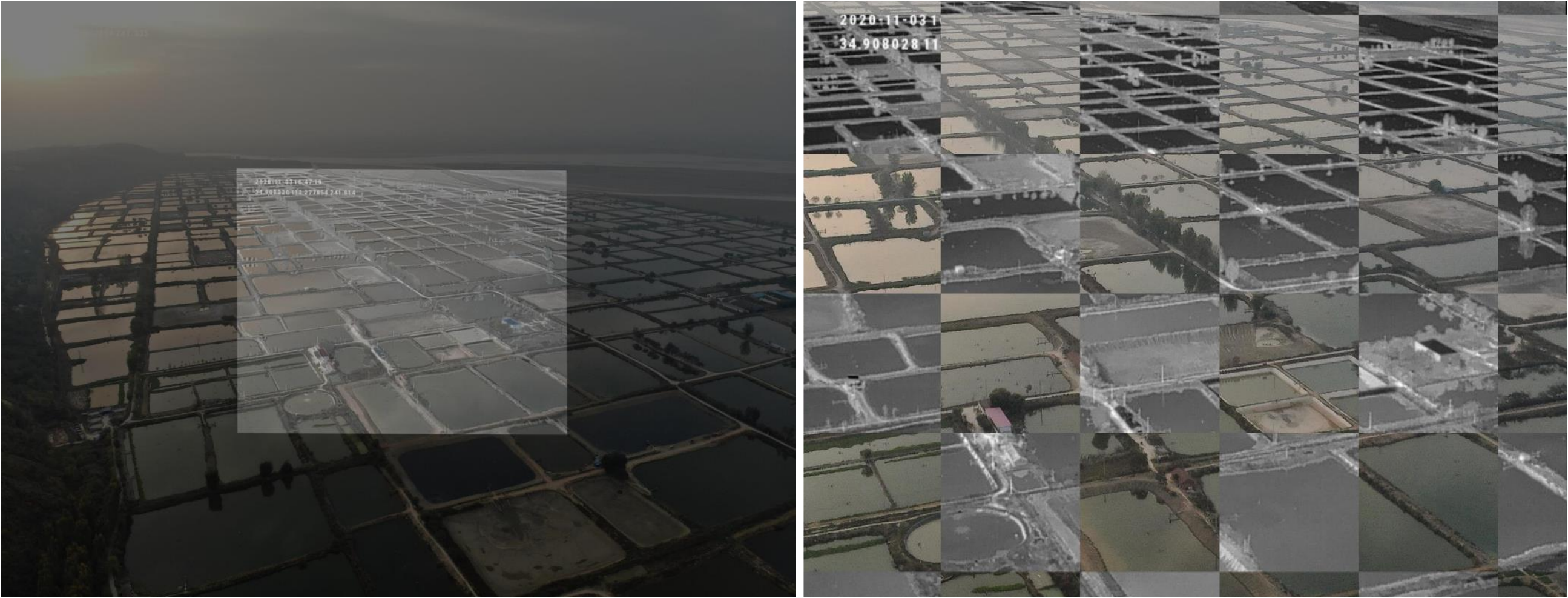}}
        \subfloat[]{\label{fig:checker:i}
        \includegraphics[height=0.8in]{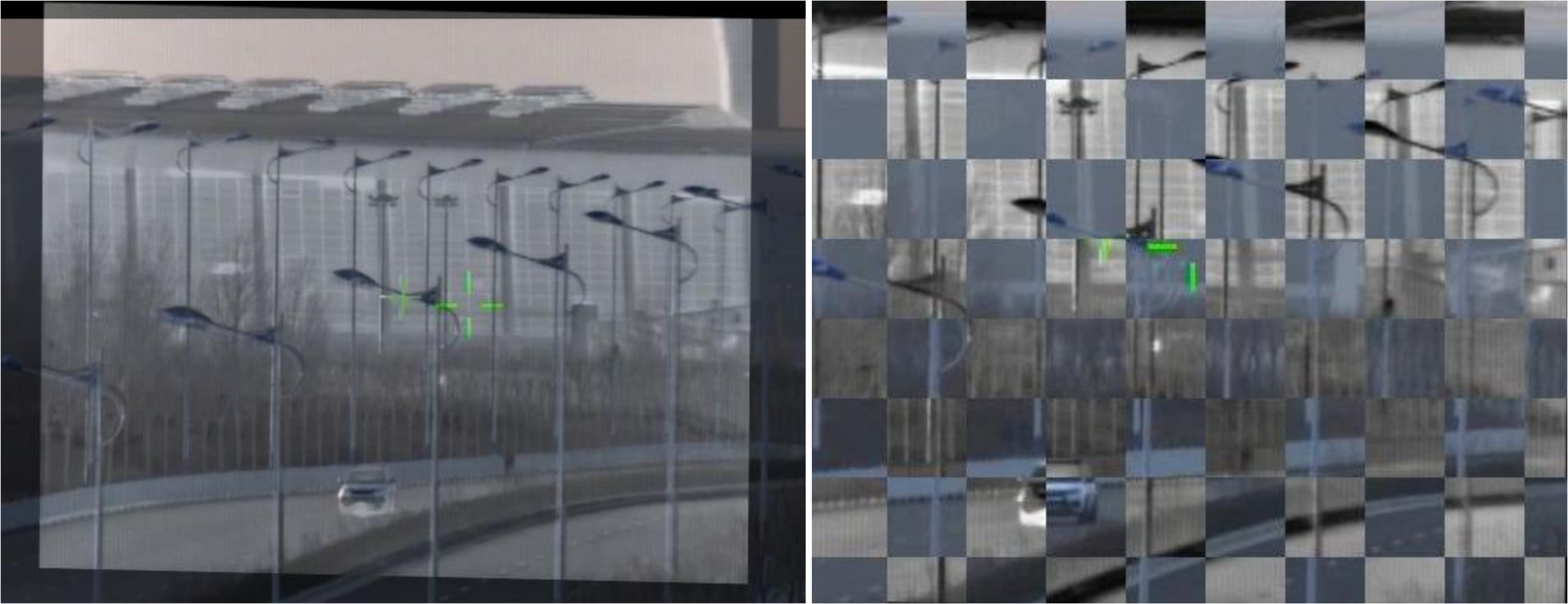}}

        \subfloat[]{\label{fig:checker:j}
        \includegraphics[height=0.8in]{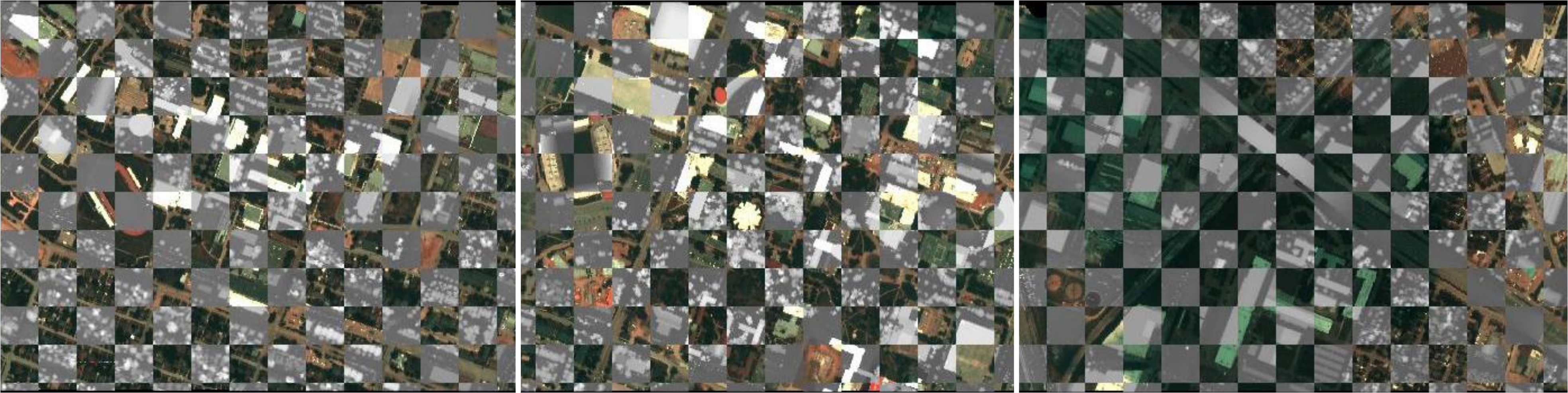}}
        \subfloat[]{\label{fig:checker:k}
        \includegraphics[height=0.8in]{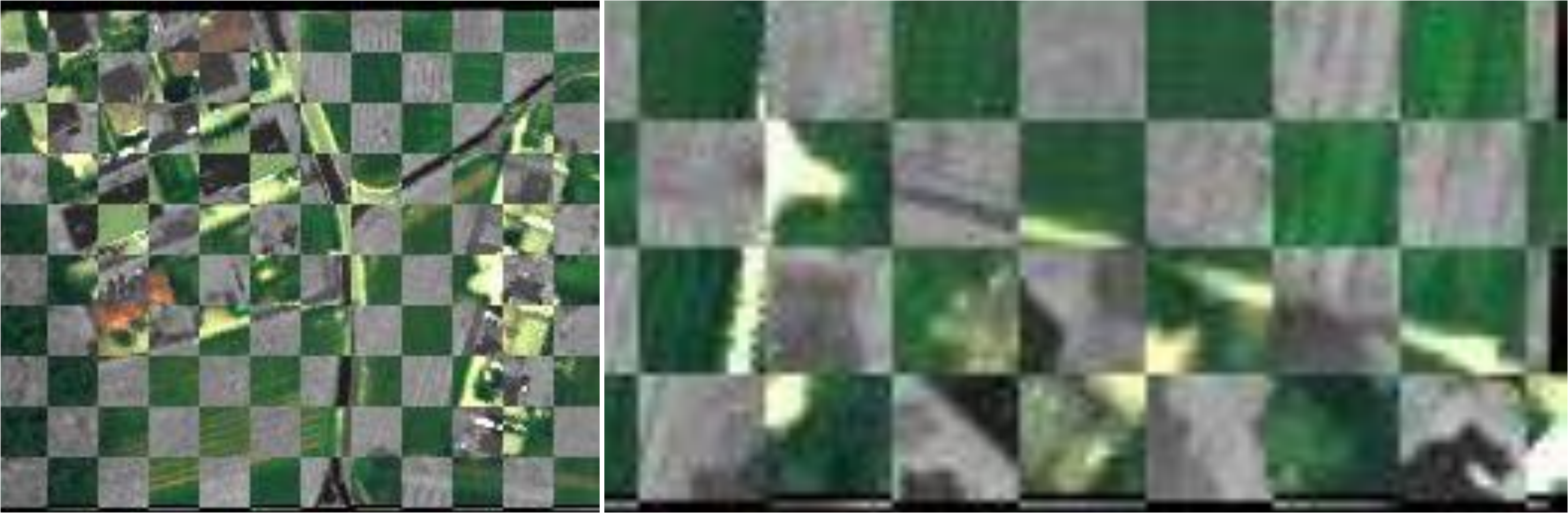}}
        \subfloat[]{\label{fig:checker:l}
        \includegraphics[height=0.8in]{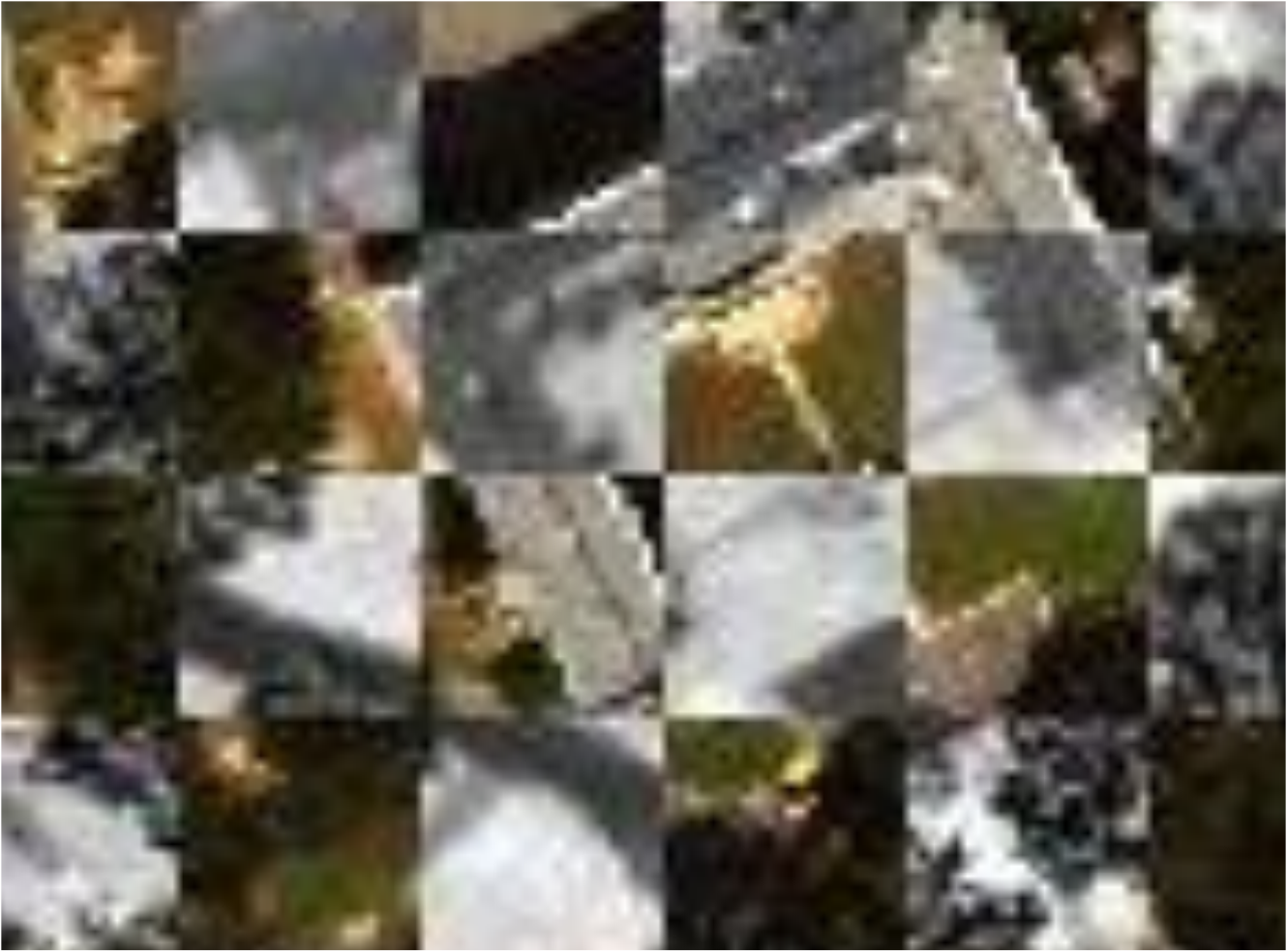}}

        \subfloat[]{\label{fig:checker:m}
        \includegraphics[height=0.85in]{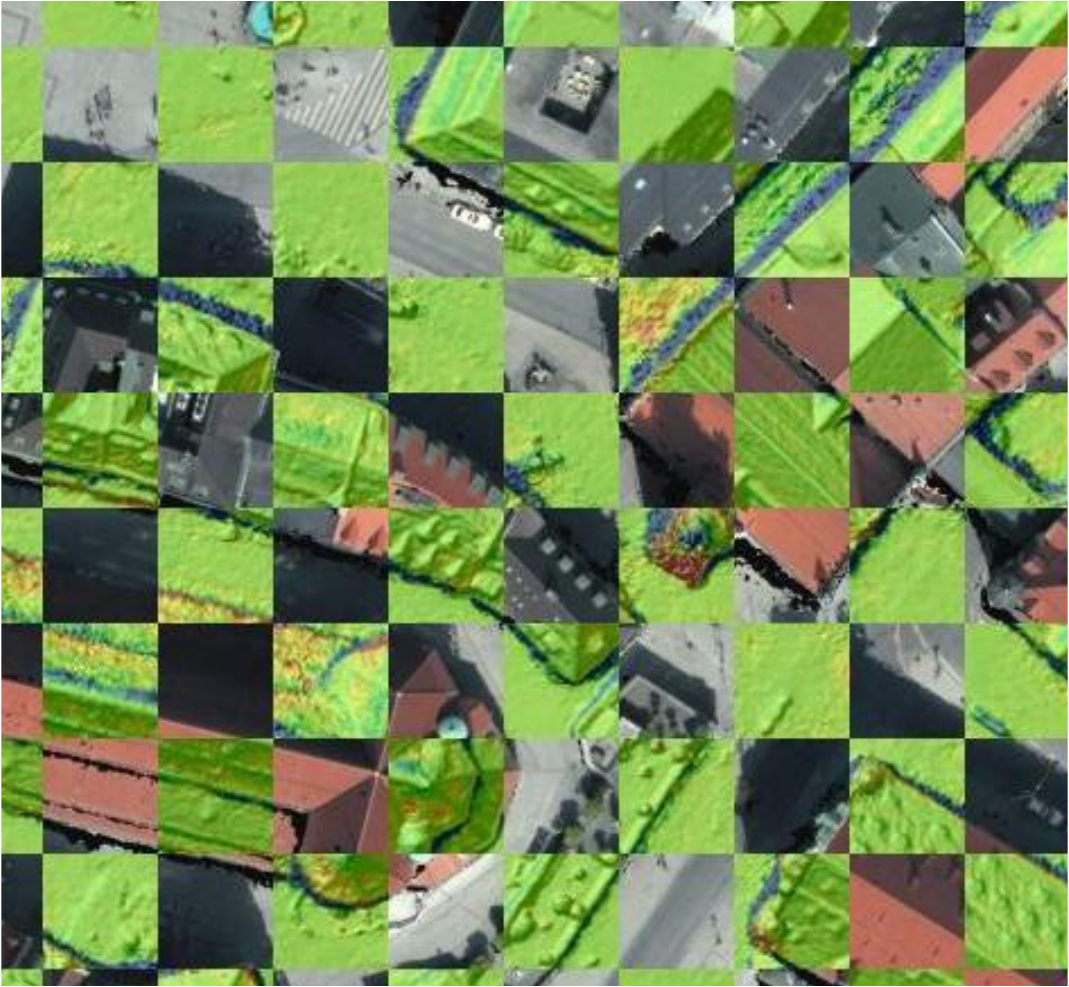}}
        \subfloat[]{\label{fig:checker:n}
        \includegraphics[height=0.9in]{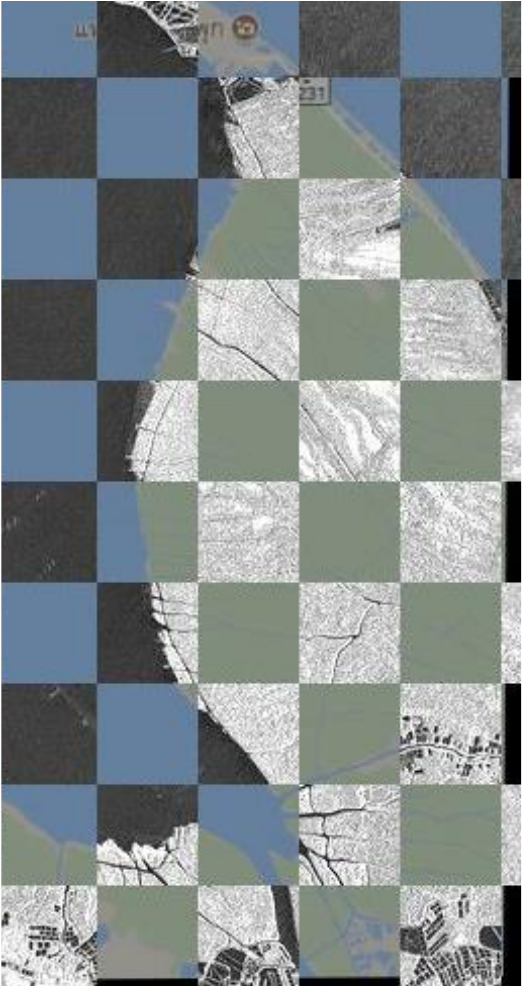}}
        \subfloat[]{\label{fig:checker:o}
        \includegraphics[height=0.85in]{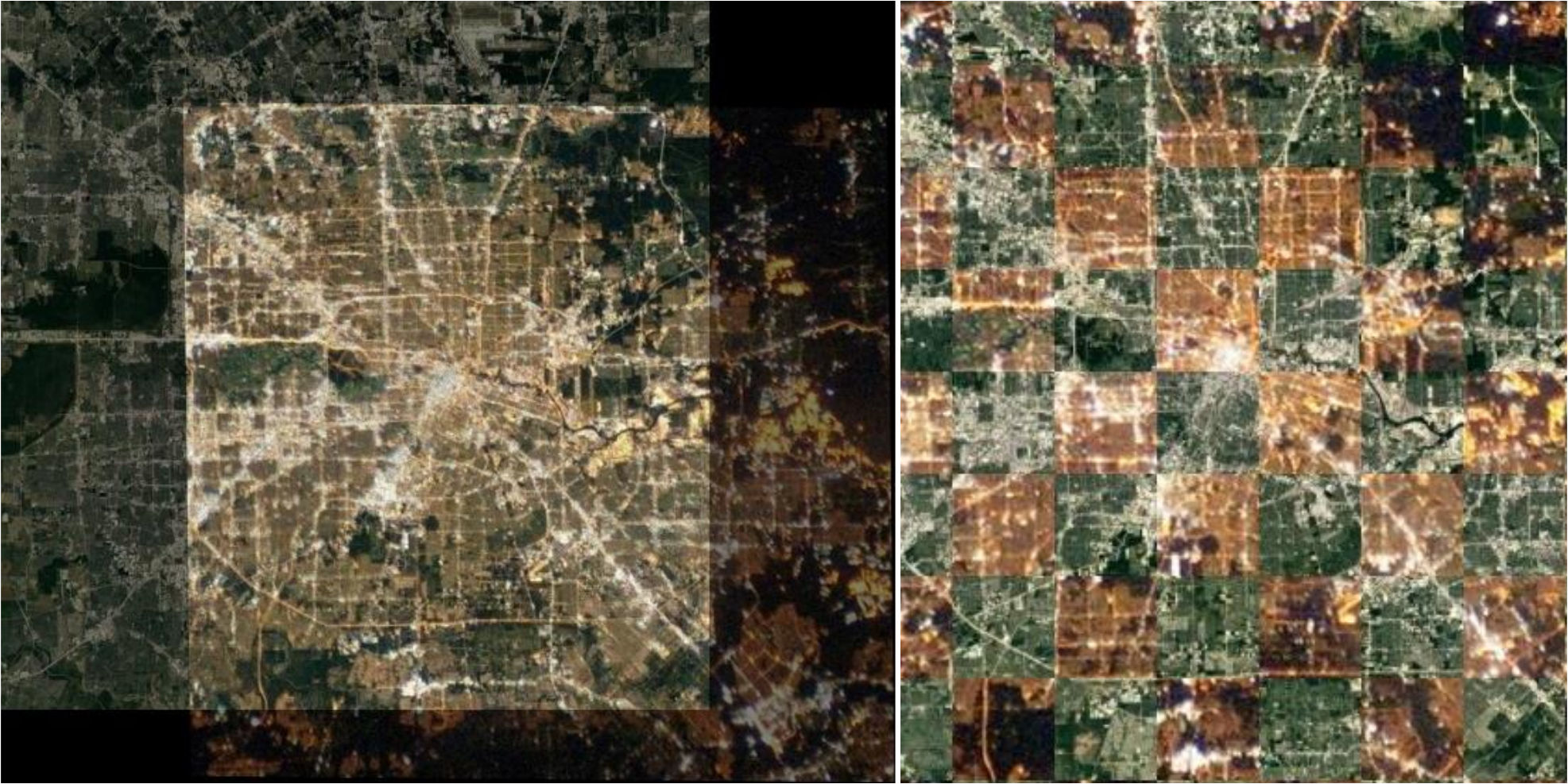}}
        \subfloat[]{\label{fig:checker:p}
        \includegraphics[height=0.85in]{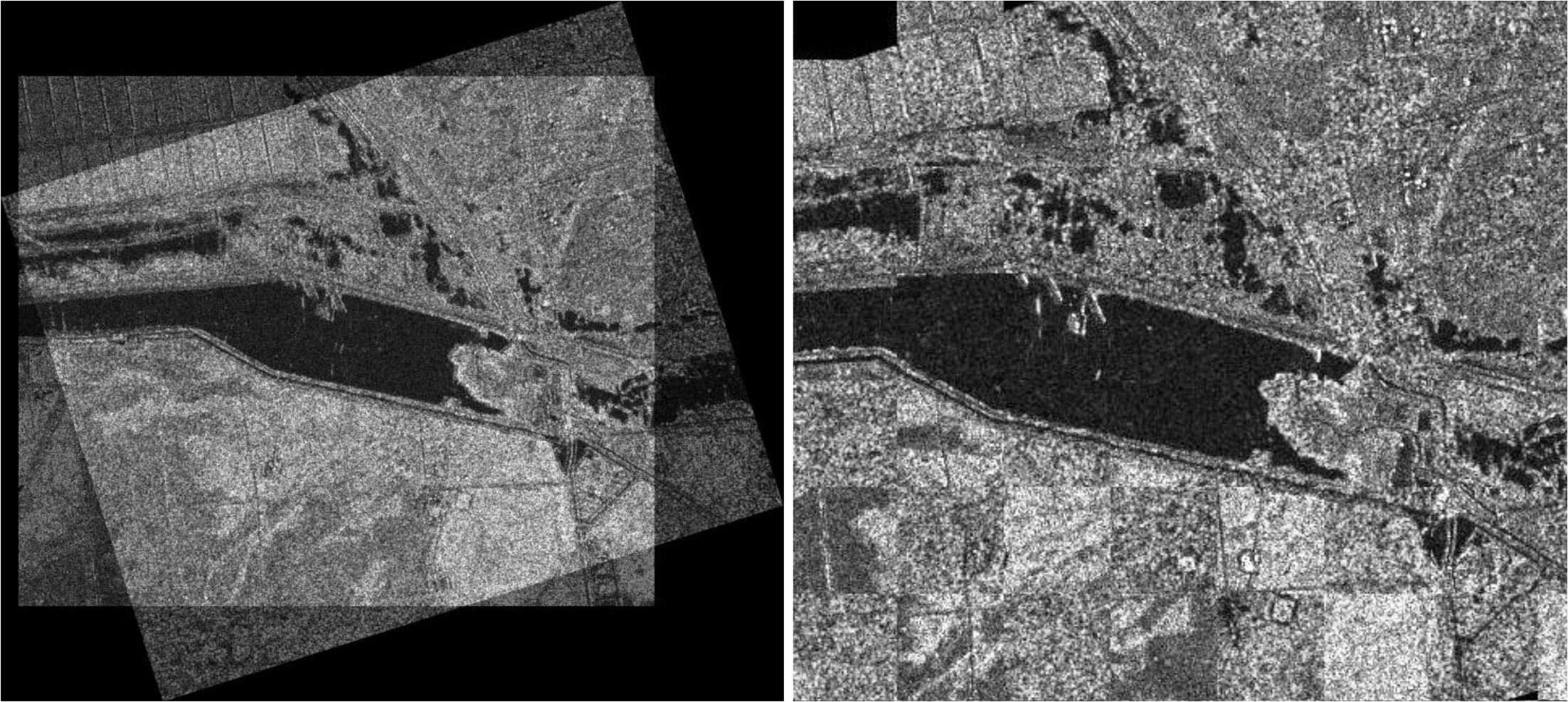}}
        \subfloat[]{\label{fig:checker:q}
        \includegraphics[height=0.85in]{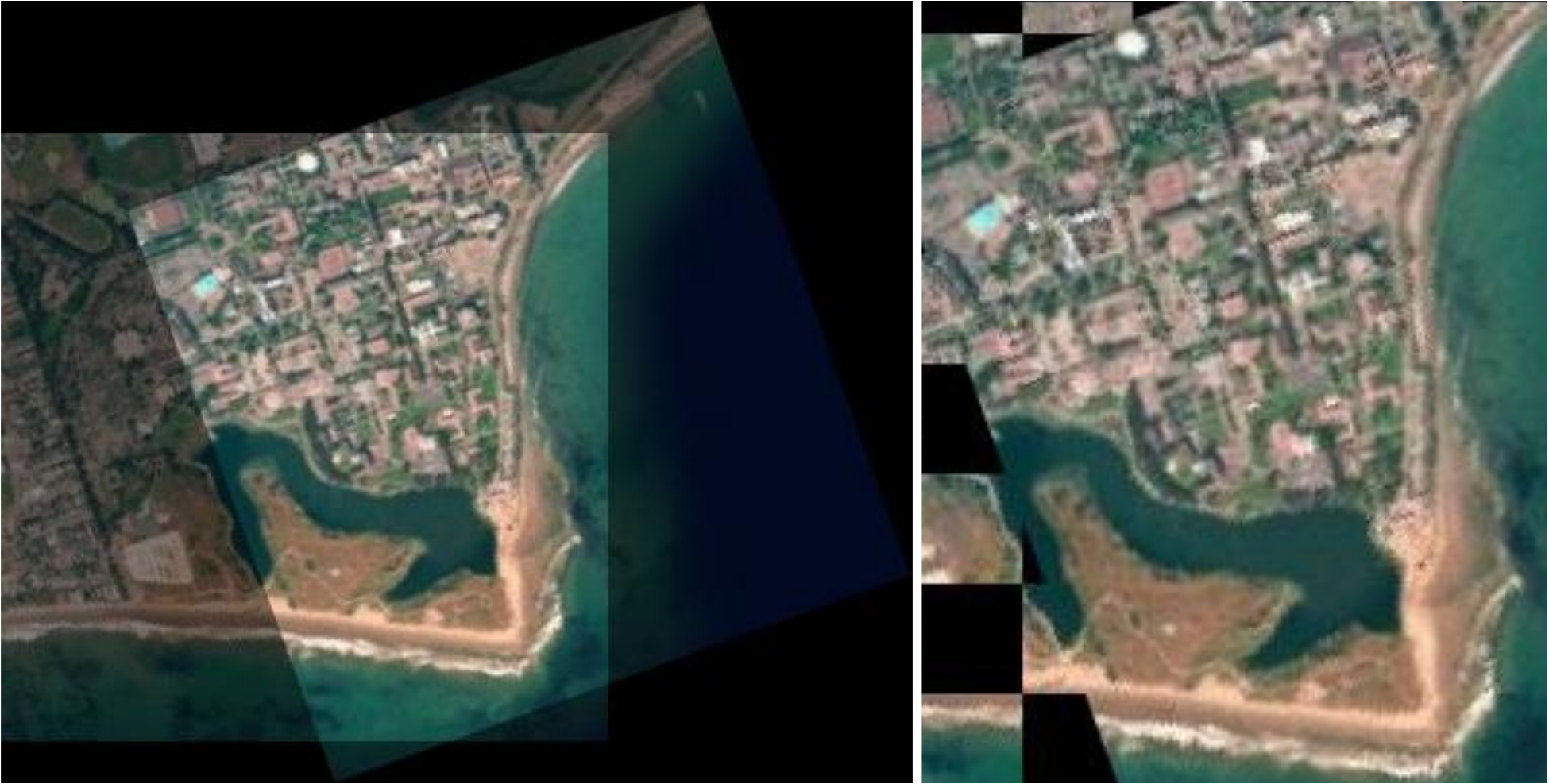}}
    \end{center}
    \caption{Registration results of the 17 remote sensing scenes by the proposed MS-HLMO, shown in fusion form and alternation form. Feature points matching results of the 17 remote sensing scenes by the proposed MS-HLMO. (a) a2:SAR-MSI, a1:MSI-HSI, a3:MSI-HSI, and a4:map-MSI. (b) MSI-HSI. (c) MSI-MSI. (d) PAN-HSI. (e) visible-infrared. (f) visible-infrared. (g) visible-infrared. (h) visible-infrared. (i) visible-infrared. (j) HSI-LiDAR. (k) HSI-LiDAR. (l) HSI-LiDAR. (m) visible-depth. (n) visible-map. (o) day-night (visible). (p) multi-source SAR. (q) single-source visible.}
    \label{fig:checker}
\end{figure*}

In this section, the registration is performed directly on each original data of the 17 scenes, to evaluate the overall registration performance. The NCMs of the eight registration methods with all 17 scenes are listed in Table \ref{tbl:result}. Fig.\ref{fig:matching} shows the corresponding visualization of feature point matching results of each scene. Due to space limitation, the one with the better result between MS-HLMO and MS-HLMO$^+$ is selected to display in each group.

In most scenes, MS-HLMO obtains the most NCM, and basically far more than other algorithms except for scene (j) and (o), in which RIFT$^+$ has more NCM. The cases in the two scenes have already been discussed in the former section. However, the proposed MS-HLMO still obtains a considerable number of correct matches that are sufficient for accurate registration, and without the interference of orientation, MS-HLMO$^+$ obtains more matches.

SIFT, SAR-SIFT, PSO-SIFT, and PIIFD fail in registration in most multi-source scenes, which cannot well cope with multi-source remote sensing data images in various situations.
Note that some of the test data selected in this paper are also used in MS-PIIFD, and the registration is successful when the number of Harris corners is set to 1000. However, when the number of Harris corners is increased to 2000 in this experiment, the effect of MS-PIIFD is greatly reduced, and it even fails. This is because the proportion of correct matching of feature points obtained by MS-PIIFD before going through FSC is low, and more outliers may be introduced after increasing the number of feature points, which makes it difficult to calculate a consistent transformation model and leads to performance degradation. It indicates that although MS-PIIFD is able to handle multi-scale and multi-mode image registration, its feature robustness and stability are limited. Under the same conditions, MS-HLMO obtains a large number of correct matches, which indicates that its feature robustness and stability are relatively high. The RIFT$^+$ works well in scenes with only intensity difference, but the NCMs are relatively low. RIFT$^+$ cannot cope with the images containing rotation and scale differences.

The last scene is a single-source image with only rotation. This scene proves that MS-HLMO can not only deal with multi-source images effectively but also has a good effect on single-source images, which still performs better than other algorithms. Among all the competitive methods, MS-HLMO is the only algorithm that does not fail in all scenes and obtains sufficient feature matches. In the scenes without obvious image rotation difference, MS-HLMO$^+$ achieves the most feature point matching, far outperforming other algorithms. Through the above experiments, the ability of MS-HLMO is verified from a detailed and comprehensive perspective, and it can effectively handle the registration task of multi-source remote sensing images. Compared with the results of the existing effective algorithms, the proposed MS-HLMO has much more significant effect than other algorithms in most cases. MS-HLMO$^+$ better deals with the situation without obvious rotation difference, which is also the majority of practical multi-source image registration tasks. MS-HLMO and its simplified version MS-HLMO$^+$ have obvious advantages in robust feature extraction and matching.



Fig.\ref{fig:checker} illustrates the registration results of the 17 scenes, in which the transformation parameters is obtained based on the feature matching shown in Fig.\ref{fig:matching}. For each image pair, one with the higher resolution is taken as the reference, and the other is aligned with it through spatial affine transformation and data interpolation. Image pairs with large spatial offset are shown in a fusion form, and the areas with prominent structures in the overlapping parts are selected and shown in an alternation form. All the operations are performed autonomously without human intervention.

It is observed that most of the image pairs are well registered, the outlines and textures of the ground covers stay continuous and consistent. The deviations are mostly within one pixel. The only obvious deviations are found in some areas of scene (g) and (i) shown in Fig.\ref{fig:checker}. This phenomenon does not result from inaccurate feature extraction or matching but is caused by the characteristics of the data. Scene (g) and (i) are ground-based images with relatively close imaging distance, in which there are obvious viewpoint differences. In addition, nonrigid geometric distortions exist in the infrared images, which may be caused by the imaging capability of the sensors. Due to the above reasons, more complex spatial differences exist between the image pairs, which is difficult to be solved with only an affine model globally transforming the images. In practice, this problem can be solved by using more sophisticate nonlinear transformation models, or by dividing the image into blocks for adaptive local registration. Apart from that, most areas is well registered, where the objects are accurately aligned.

\section{Conclusions}
\label{sec:conclusions}
In this paper, an image registration algorithm has been proposed for multi-source remote sensing image registration. The proposed MS-HLMO utilizes Partial Main Orientation Map (PMOM) based on Average Squared Gradient (ASG) as the basic feature map, a generalized GLOH-like (GGLOH) descriptor structure, and multi-scale feature extraction and matching strategy, which has the benefit of high robustness to multimodal properties and effectively deals with the key problems of intensity distortion, rotation, and scale differences in registration of multi-source images. Comprehensive experiments on 17 datasets of multi-source remote sensing scenes reveal the effectiveness and generalization of the proposed MS-HLMO and its simplified version MS-HLMO$^+$ compared with other state-of-the-art registration methods. 

\vspace{11pt}


\vfill

\bibliographystyle{IEEEtran}
\bibliography{refs}

\end{document}